\documentclass[12pt]{article}
\usepackage{a4p}
\usepackage{epsfig}
\usepackage{array}
\usepackage{cite}
\usepackage{pennames}
\usepackage{rotating}
\newcommand{\PPEnum} {CERN-EP/98-108}

\newcommand{\Date}      {3rd July 1998}
\newcommand{\inmath}[1] {\ifmmode#1\else$#1$\fi}
\newcommand{\definmath}[2] {\def#1{\ifmmode#2\else$#2$\fi}}
\newcommand{\alphas} {\alpha_{\mathrm{s}}}
\newcommand{\alphaem} {\mbox{$\alpha_{\mathrm{em}}$}}

\definmath{\PWpm} {\mathrm{W}^{\pm}}      
\definmath{\Plp} {\ell^{+}}        
\definmath{\Plm} {\ell^{-}}        
\definmath{\Plpm}   {\ell^{\pm}}         
\definmath{\Pgtp} {\tau^{+}}        
\definmath{\Pgtm} {\tau^{-}}        
\definmath{\Pgtpm}   {\tau^{\pm}}         
\definmath{\Pgn}  {\nu}          
\definmath{\Pagn} {\overline{\nu}}     
\definmath{\Pf}      {\mathrm{f}}
\definmath{\Paf}  {\overline{\mathrm{f}}}
\definmath{\Pq}      {\mathrm{q}}
\definmath{\Paq}  {\overline{\mathrm{q}}}
\definmath{\Pu}      {\mathrm{u}}
\definmath{\Pau}  {\overline{\mathrm{u}}}
\definmath{\Pd}      {\mathrm{d}}
\definmath{\Pad}  {\overline{\mathrm{d}}}
\definmath{\Ps}      {\mathrm{s}}
\definmath{\Pas}  {\overline{\mathrm{s}}}
\definmath{\Pc}      {\mathrm{c}}
\definmath{\Pac}  {\overline{\mathrm{c}}}
\definmath{\Pb}      {\mathrm{b}}
\definmath{\Pab}  {\overline{\mathrm{b}}}
\definmath{\Pt}      {\mathrm{t}}
\definmath{\Pat}  {\overline{\mathrm{t}}}
\definmath{\Pap}  {\overline{\mathrm{p}}}
\definmath{\Pan}  {\overline{\mathrm{n}}}
\definmath{\PaD}  {\overline{\mathrm{D}}}
\definmath{\PaDz} {\overline{\mathrm{D}}^{0}}
\definmath{\PaB}  {\overline{\mathrm{B}}}
\definmath{\PaBz} {\overline{\mathrm{B}}^{0}}
\definmath{\PsDpm}   {\mathrm{D}^{\pm}_{\mathrm{s}}}  
\definmath{\PcgLpm}  {\Lambda^{\pm}_{\mathrm{c}}}  
\definmath{\PD} {\mathrm{D}}     
\definmath{\PDst} {\mathrm{D}^{*}}     
\definmath{\PgLz} {\Lambda^{0}}        

\newcommand{\snu}{\tilde{\nu}}

\newcommand{\massof}[1] {m_{\smash{#1}\mathstrut}}
\newcommand{\mtop}   {\massof{\mathrm{top}}}
\newcommand{\mHiggs} {\massof{\mathrm{Higgs}}}

\newcommand{\mPZ} {\massof{\mathrm{Z}}}
\newcommand{\mX}  {\massof{\mathrm{X}}}
\newcommand{\Rb}  {\mbox{$R_{\rm b}$}}

\newcommand{\AFB}    {A_{\mathrm{FB}}}
\newcommand{\AFBSM}  {A_{\mathrm{FB}}^{\mathrm{SM}}}

\newcommand{\thacol}{\theta_{\mathrm{acol}}}

\newcommand{\epem}   {\Pep\Pem}

\newcommand{\mumu}   {\Pgmp\Pgmm}
\newcommand{\tautau} {\Pgtp\Pgtm}
\newcommand{\ff}     {\Pf\Paf}

\newcommand{\qqbar}  {\Pq\Paq}
\newcommand{\uubar}  {\Pu\Pau}
\newcommand{\ddbar}  {\Pd\Pad}

\newcommand{\bbbar}  {\Pb\Pab}

\newcommand{\ud}     {\uubar + \ddbar}

\newcommand{\WW}{\ensuremath{\mathrm{W}^+\mathrm{W}^-}}
\newcommand{\eetoee}    {\epem\to\epem}

\newcommand{\eetomumu}     {\epem\to\mumu}
\newcommand{\eetotautau}   {\epem\to\tautau}

\newcommand{\eetoqq}    {\epem\to\qqbar}

\newcommand{\dsdcc}        {{\rm d}\sigma/{\rm d}\!\cos\theta}
\newcommand{\dsdabscc}     {{\rm d}\sigma/{\rm d}|\cos\theta|}
\def\lapproxeq {\mbox{{\lower .7ex\hbox{$\;\stackrel{\textstyle
                  <}{\sim}\;$}}}}
\def\gapproxeq  {\mbox{{\lower .7ex\hbox{$\;\stackrel{\textstyle
                  >}{\sim}\;$}}}}
\newcommand{\roots} {\sqrt{s}}

\newcommand {\ct}      {\mbox{$\cos \theta$}}
\newcommand {\absct}   {\mbox{$|\cos \theta |$}}

\newcommand {\absctem} {\mbox{$|\cos \theta_{{\rm e}^-} |$}}

\newcommand {\absctep} {\mbox{$|\cos \theta_{{\rm e}^+} |$}}

\definmath{\GeV}  {\mathrm{GeV}}
\definmath{\GeVc} {\mathrm{GeV}\!/c}
\definmath{\GeVcc}   {\mathrm{GeV}\!/c^2}
\definmath{\MeV}  {\mathrm{MeV}}
\definmath{\MeVc} {\mathrm{MeV}\!/c}
\definmath{\MeVcc}   {\mathrm{MeV}\!/c^2}
\definmath{\MVm}  {\mathrm{MV}\!/\mathrm{m}}
\definmath{\keV}  {\mathrm{keV}}
\definmath{\keVcm}   {\mathrm{keV}\!/\mathrm{cm}}
\definmath{\kV}      {\mathrm{kV}}
\definmath{\km}      {\mathrm{km}}
\definmath{\meter}   {\mathrm{m}}
\definmath{\cm}      {\mathrm{cm}}
\definmath{\mm}      {\mathrm{mm}}
\definmath{\micron}  {\mu\mathrm{m}}
\definmath{\nm}      {\mathrm{nm}}
\definmath{\kg}      {\mathrm{kg}}
\definmath{\gram} {\mathrm{g}}
\definmath{\second}  {\mathrm{s}}
\definmath{\microsec}   {\mu\mathrm{s}}
\definmath{\degree}  {^\circ}
\definmath{\degC} {^\circ\mathrm{C}}
\definmath{\ohm}  {\Omega}
\definmath{\Mohm} {\mathrm{M}\Omega}
\definmath{\rad}  {\mathrm{rad}}
\definmath{\mrad} {\mathrm{mrad}}
\definmath{\nb}      {\mathrm{nb}}
\newcommand{\eqref}[1]  {(\ref{#1})}
\newcommand{\PhysLett}  {Phys.~Lett.}
\newcommand{\PRL} {Phys.~Rev.\ Lett.}
\newcommand{\PhysRep}   {Phys.~Rep.}
\newcommand{\PhysRev}   {Phys.~Rev.}
\newcommand{\NPhys}  {Nucl.~Phys.}
\newcommand{\NIM} {Nucl.~Instr.\ Meth.}
\newcommand{\ZPhys}  {Z.~Phys.}
\newcommand{\IEEENS} {IEEE Trans.\ Nucl.~Sci.}
\newcommand{\CPC} {Comput. Phys. Commun.}
\newcommand{\EPJ} {Eur.~Phys.~J.} 
\newcommand{\ALEPHColl}   {ALEPH Collab.}

\newcommand{\LthreeColl}  {L3 Collab.}
\newcommand{\OPALColl}    {OPAL Collab.}
\newcolumntype{L} {>{$}l<{$}}
\newcolumntype{C} {>{$}c<{$}}
\newcolumntype{R} {>{$}r<{$}}

   \newcommand{\lept}
{\ell^+\ell^-}   

\newcommand{\epsz}  {\varepsilon_0}   \newcommand{\lamm}   {\Lambda_-}
\newcommand{\lamp} {\Lambda_+} 

\begin{document}
%
%
\begin{titlepage}
%
\begin{center}
    \Large EUROPEAN LABORATORY FOR PARTICLE PHYSICS
\end{center}
\bigskip 
\begin{flushright}
    \large \PPEnum\\ \Date \\
\end{flushright}
%
%
\begin{center}
    \huge\bf\boldmath Tests of  the Standard Model and Constraints  on
    New  Physics  from  Measurements of Fermion-pair  Production  at
    183~\GeV\ at LEP 
\end{center}
\vspace{0.5cm} 
%
%
\begin{center}
    \LARGE   The  OPAL   Collaboration   \\  
\vspace{0.5cm} 
%
%
\begin{abstract}
Cross-sections for hadronic, \bbbar\ and lepton pair final states in 
\epem\ collisions at $\sqrt{s}$=183~GeV, measured with the OPAL detector 
at LEP, are presented and compared with the predictions of the Standard
Model. Forward-backward asymmetries for the leptonic final states have
also been measured. Cross-sections and asymmetries are also presented for 
data recorded in 1997 at $\sqrt{s}$=130 and 136~GeV. The results are used 
to measure the energy dependence of the electromagnetic coupling constant 
\alphaem, and to place limits on new physics as described by four-fermion 
contact interactions or by the exchange of a new heavy particle such as a 
leptoquark, or of a squark or sneutrino in supersymmetric theories with 
$R$-parity violation. 
\end{abstract}
\end{center}
\vspace{0.5cm}
\begin{center}
{\large Submitted to \EPJ\ C}
\end{center}
%
%
\end{titlepage}

\begin{center}{\Large        The OPAL Collaboration
}\end{center}\bigskip
\begin{center}{
G.\thinspace Abbiendi$^{  2}$,
K.\thinspace Ackerstaff$^{  8}$,
G.\thinspace Alexander$^{ 23}$,
J.\thinspace Allison$^{ 16}$,
N.\thinspace Altekamp$^{  5}$,
K.J.\thinspace Anderson$^{  9}$,
S.\thinspace Anderson$^{ 12}$,
S.\thinspace Arcelli$^{ 17}$,
S.\thinspace Asai$^{ 24}$,
S.F.\thinspace Ashby$^{  1}$
% ...
}\end{center}\bigskip
\bigskip
$^{  1}$School of Physics and Astronomy, University of Birmingham,
Birmingham B15 2TT, UK
\newline
$^{  2}$Dipartimento di Fisica dell' Universit\`a di Bologna and INFN,
I-40126 Bologna, Italy
\newline
$^{  3}$Physikalisches Institut, Universit\"at Bonn,
D-53115 Bonn, Germany
\newline
$^{  4}$Department of Physics, University of California,
Riverside CA 92521, USA
\newline
$^{  5}$Cavendish Laboratory, Cambridge CB3 0HE, UK
\newline
$^{  6}$Ottawa-Carleton Institute for Physics,
Department of Physics, Carleton University,
Ottawa, Ontario K1S 5B6, Canada
\newline
$^{  7}$Centre for Research in Particle Physics,
Carleton University, Ottawa, Ontario K1S 5B6, Canada
\newline
$^{  8}$CERN, European Organisation for Particle Physics,
CH-1211 Geneva 23, Switzerland
\newline
$^{  9}$Enrico Fermi Institute and Department of Physics,
University of Chicago, Chicago IL 60637, USA
\newline
$^{ 10}$Fakult\"at f\"ur Physik, Albert Ludwigs Universit\"at,
D-79104 Freiburg, Germany
\newline
$^{ 11}$Physikalisches Institut, Universit\"at
Heidelberg, D-69120 Heidelberg, Germany
\newline
$^{ 12}$Indiana University, Department of Physics,
Swain Hall West 117, Bloomington IN 47405, USA
\newline
$^{ 13}$Queen Mary and Westfield College, University of London,
London E1 4NS, UK
\newline
$^{ 14}$Technische Hochschule Aachen, III Physikalisches Institut,
Sommerfeldstrasse 26-28, D-52056 Aachen, Germany
\newline
$^{ 15}$University College London, London WC1E 6BT, UK
\newline
$^{ 16}$Department of Physics, Schuster Laboratory, The University,
Manchester M13 9PL, UK
\newline
$^{ 17}$Department of Physics, University of Maryland,
College Park, MD 20742, USA
\newline
$^{ 18}$Laboratoire de Physique Nucl\'eaire, Universit\'e de Montr\'eal,
Montr\'eal, Quebec H3C 3J7, Canada
\newline
$^{ 19}$University of Oregon, Department of Physics, Eugene
OR 97403, USA
\newline
$^{ 20}$CLRC Rutherford Appleton Laboratory, Chilton,
Didcot, Oxfordshire OX11 0QX, UK
\newline
$^{ 22}$Department of Physics, Technion-Israel Institute of
Technology, Haifa 32000, Israel
\newline
$^{ 23}$Department of Physics and Astronomy, Tel Aviv University,
Tel Aviv 69978, Israel
\newline
$^{ 24}$International Centre for Elementary Particle Physics and
Department of Physics, University of Tokyo, Tokyo 113, and
Kobe University, Kobe 657, Japan
\newline
$^{ 25}$Institute of Physical and Environmental Sciences,
Brunel University, Uxbridge, Middlesex UB8 3PH, UK
\newline
$^{ 26}$Particle Physics Department, Weizmann Institute of Science,
Rehovot 76100, Israel
\newline
$^{ 27}$Universit\"at Hamburg/DESY, II Institut f\"ur Experimental
Physik, Notkestrasse 85, D-22607 Hamburg, Germany
\newline
$^{ 28}$University of Victoria, Department of Physics, P O Box 3055,
Victoria BC V8W 3P6, Canada
\newline
$^{ 29}$University of British Columbia, Department of Physics,
Vancouver BC V6T 1Z1, Canada
\newline
$^{ 30}$University of Alberta,  Department of Physics,
Edmonton AB T6G 2J1, Canada
\newline
$^{ 31}$Duke University, Dept of Physics,
Durham, NC 27708-0305, USA
\newline
$^{ 32}$Research Institute for Particle and Nuclear Physics,
H-1525 Budapest, P O  Box 49, Hungary
\newline
$^{ 33}$Institute of Nuclear Research,
H-4001 Debrecen, P O  Box 51, Hungary
\newline
$^{ 34}$Ludwigs-Maximilians-Universit\"at M\"unchen,
Sektion Physik, Am Coulombwall 1, D-85748 Garching, Germany
\newline
\bigskip\newline
$^{  a}$ and at TRIUMF, Vancouver, Canada V6T 2A3
\newline
$^{  b}$ and Royal Society University Research Fellow
\newline
$^{  c}$ and Institute of Nuclear Research, Debrecen, Hungary
\newline
$^{  d}$ and Department of Experimental Physics, Lajos Kossuth
University, Debrecen, Hungary
\newline
$^{  e}$ on leave of absence from the University of Freiburg
\newline
\newpage
 
\section{Introduction}           \label{sec:intro}
The LEP accelerator has provided \epem\
collisions at ever increasing energies over the past two years.
In this paper we present measurements of cross-sections for hadronic,
\bbbar\ and lepton pair final states at a centre-of-mass energy $\sqrt{s}$
of 183~GeV; forward-backward
asymmetries for the leptonic states are also given. The data
were collected by the OPAL detector at LEP in 1997. During 1997 LEP
briefly returned to centre-of-mass energies of 130 and 136~GeV,
providing an integrated luminosity at these energies similar to
the 1995 `LEP1.5' run. We also present cross-sections and asymmetry
measurements from these data, and combine them with the results from the 
earlier run~\cite{bib:OPAL-SM172}.

The analyses presented here are essentially the same as those already
presented at lower energies~\cite{bib:OPAL-SM172}. We use identical
techniques to measure $s'$, the square of the centre-of-mass
energy of the \epem\ system after initial-state radiation, and 
to separate `non-radiative' events, which have little initial-state
radiation, from `radiative return' to the Z peak. However, we have changed 
the definition of non-radiative events compared with~\cite{bib:OPAL-SM172}, 
loosening the requirement from $s'/s > 0.8$ to $s'/s > 0.7225$. This new
definition reduces the total measurement errors at the highest centre-of-mass
energies, and is the 
recommended definition of the LEP Electroweak Working Group. Inclusive
measurements are corrected to $s'/s > 0.01$ as before. As in our previous
publication, we correct our measurements of hadronic, \mumu\ and \tautau\
events for the effect of interference between initial- and final-state 
radiation. We use the same treatment as before of the four-fermion contribution 
to the two-fermion final states. Because of ambiguities arising from
the $t$-channel contribution, for the \epem\ final state the acceptance
is defined in terms of the angle $\theta$ of the electron or positron
with respect to the electron beam direction and the acollinearity angle
$\thacol$ between the electron and positron. Cross-sections
and asymmetries for \epem\ are not corrected for interference between initial-
and final-state radiation; they are compared to theoretical predictions
which include interference.

Measurements of fermion-pair production up to 172~GeV have shown very 
good agreement with Standard Model
expectations~\cite{bib:OPAL-SM172,bib:ADL-SM}. Here we repeat our
measurement of the electromagnetic coupling constant \alphaem($\sqrt{s}$)
including the higher energy data. Including data at 183~GeV also
allows us to extend the searches for new physics presented 
in~\cite{bib:OPAL-SM172}. In particular we obtain improved limits
on the energy scale of a possible four-fermion contact interaction.
We also present improved limits on the coupling of 
a heavy particle such as a leptoquark, or a scalar quark (squark) in 
supersymmetric theories with $R$-parity violation, which might affect the 
hadronic cross-section via a $t$-channel exchange diagram. These analyses 
are updates of those already presented in~\cite{bib:OPAL-SM172}. In this paper we
extend our search for heavy particles to include those coupling
to leptons only. Such particles could be scalar neutrinos (sneutrinos) in 
supersymmetric theories with $R$-parity violation. Searches for such particles have
been presented in~\cite{bib:L3-rpvsnu}. In this paper we introduce a new 
technique in which a scan over the complete $s'$ distribution improves our
sensitivity for masses between the centre-of-mass energy points of
LEP for processes involving $s$-channel sneutrinos.

The paper is organized as follows. In section~\ref{sec:data} we describe
the data analysis, cross-section and asymmetry measurements. As the
analyses are essentially the same as in~\cite{bib:OPAL-SM172} we give
only a brief description of any changes. In section~\ref{sec:sm} we
compare our measurements to the predictions of the Standard Model
and use them to measure the energy dependence of $\alphaem$. The results of 
searches for new physics are presented in section~\ref{sec:new_phys}.

\section{Data Analysis}           \label{sec:data}
The OPAL detector\footnote{OPAL uses a right-handed coordinate system in
which the $z$ axis is along the electron beam direction and the $x$
axis is horizontal. The polar angle $\theta$ is measured with respect
to the $z$ axis and the azimuthal angle $\phi$ with respect to the
$x$ axis.}, trigger and data acquisition system are fully described 
elsewhere~\cite{bib:OPAL-detector,bib:OPAL-SI,bib:OPAL-SW,bib:OPAL-TR,
bib:OPAL-DAQ}. The analyses presented in this paper use 54--57~pb$^{-1}$ of 
data collected at centre-of-mass energies of 181--184~GeV during 1997; 
the actual amount of data varies from channel to channel. The
luminosity-weighted mean centre-of-mass energy is
182.69$\pm$0.06~GeV~\cite{bib:ELEP}. In addition, we present results for
about 2.6 (3.3)~pb$^{-1}$ of data at 130.00$\pm$0.03 (135.98$\pm$0.03)~GeV 
collected during 1997 and combine these with earlier results at 
similar centre-of-mass energies~\cite{bib:OPAL-SM172}. 

Selection efficiencies and backgrounds were calculated using Monte
Carlo simulations. The set of generators used is identical to that
in~\cite{bib:OPAL-SM172}. At 183~GeV a new background arises from
the production of Z-pair events; these were simulated with the
grc4f~\cite{bib:grc4f} and PYTHIA~\cite{bib:pythia} generators.
All events were passed through a simulation~\cite{bib:gopal} of the OPAL 
detector and processed as for real data.

The luminosity was measured using small-angle Bhabha scattering events
recorded in the silicon-tungsten luminometer~\cite{bib:OPAL-SW,bib:OPAL-SM172}.
The overall error on the luminosity measurement amounts to 0.43\%
at 183~GeV, arising mainly from data statistics (0.26\%) and knowledge
of the theoretical cross-section (0.25\%). At 130 (136)~GeV the total
error of 1.0\% (1.0\%) arises mainly from data statistics. Errors from
the luminosity measurement are included in all the systematic errors
on cross-sections quoted in this paper, and correlations between
measurements arising from the luminosity determination are included
in all fits.

\subsection{\bf Measurements at $\protect\sqrt{s}$ = 183~GeV}
Hadronic, \epem, \mumu\ and \tautau\ events were selected using the
criteria described in~\cite{bib:OPAL-SM172} with some modifications
to improve efficiency and background at the higher centre-of-mass
energy. Here we briefly describe these changes.
\begin{itemize}
 \item In the selection of \qqbar\ events we have rejected events 
       identified as W-pairs according to the
       criteria of~\cite{bib:OPAL-WW172}, instead of subtracting
       their expected contribution, resulting in a reduction of about 15\%
       in the overall error on the non-radiative cross-section. 
 \item The background in the muon pair sample has been reduced 
       from 11.5\% to 4.7\% for inclusive events, from 6.7\% to 2.0\%
       for non-radiative events, by
       introducing a cut on the invariant mass of the muon pair.
       For inclusive events, if the ratio of the visible 
       energy\footnote{The visible energy is defined as the scalar sum of 
       the momenta of the two muons plus the energy of the highest energy 
       cluster in the electromagnetic calorimeter~\cite{bib:OPAL-SM172}.}
       to the centre-of-mass energy is less than 
       $0.5 (\mPZ^{2} / s) + 0.75$ the muon pair invariant
       mass is required to be greater than 70~GeV. For non-radiative
       events the mass is required to be greater than
       $\sqrt{(\mPZ^{2} + 0.1 s)}$ (about 108~GeV).
 \item The efficiency for tau pairs, particularly radiative events,
       has been increased by extending the angular acceptance and
       adjusting cuts so that additional background is suppressed.
       The result of the changes is to increase the efficiency for inclusive 
       events from 31\% to 40\%, with a modest (2\%) increase in background.
       In detail the changes are:
 \begin{itemize}
  \item The acceptance has been increased by requiring that the
        value of $\absct$ for both tau leptons satisfies $\absct < 0.9$, 
        instead of demanding that the average value satisfy
        $|\ct_{\mathrm{av}}| < 0.85$.
  \item The cuts on the total energy of an event have been modified:
        as before, the total visible energy in the event, derived from 
        the scalar sum of all track momenta plus electromagnetic calorimeter 
        energy, was required to be less than 1.1$\roots$, but the lower
        cut on this variable was removed. Instead, the total electromagnetic 
        calorimeter energy was required to be greater than 0.02$\roots$ and 
        either the total electromagnetic calorimeter energy or the scalar 
        sum of track momenta was required to be greater than 0.2$\roots$.
  \item The cuts on the missing momentum and its direction, calculated
        using electromagnetic calorimeter clusters, have been modified:
        the missing momentum in the plane transverse to the beam
        axis was required to exceed 0.015$\roots$, and the polar angle
        of the missing momentum was required to satisfy $\absct < 0.99$.
        The cut on the polar angle of the missing momentum calculated
        using central detector tracks was removed.
  \item The above modifications increase the efficiency substantially,
        but also increase the background, particularly from Bhabha events. 
        To reduce this background two new cuts have been introduced. 
        Using the values of $\theta$ measured for the two tau leptons
        (as defined in~\cite{bib:OPAL-SM172}),
        the expected energy of each lepton is calculated assuming that
        the final state consists only of two leptons plus a single
        unobserved photon along the beam direction. We then require
        that
        \[ 0.02 < \sqrt{(X_{E1}^2 + X_{E2}^2)} < 0.8,
        \]
        and
        \[ \sqrt{(X_{P1}^2 + X_{P2}^2)} < 0.8,
        \]
        where $X_{E1,E2}$ are the total electromagnetic calorimeter
        energies in each tau cone normalized to the expected value
        calculated above, and $X_{P1,P2}$ are the scalar sums of track
        momenta in the two tau cones, also normalized to the
        expected values. These cuts are designed to remove both electron
        and muon pairs.
 \end{itemize}
\end{itemize}

\begin{figure}
  \epsfxsize=\textwidth
  \epsfbox[0 0 567 567]{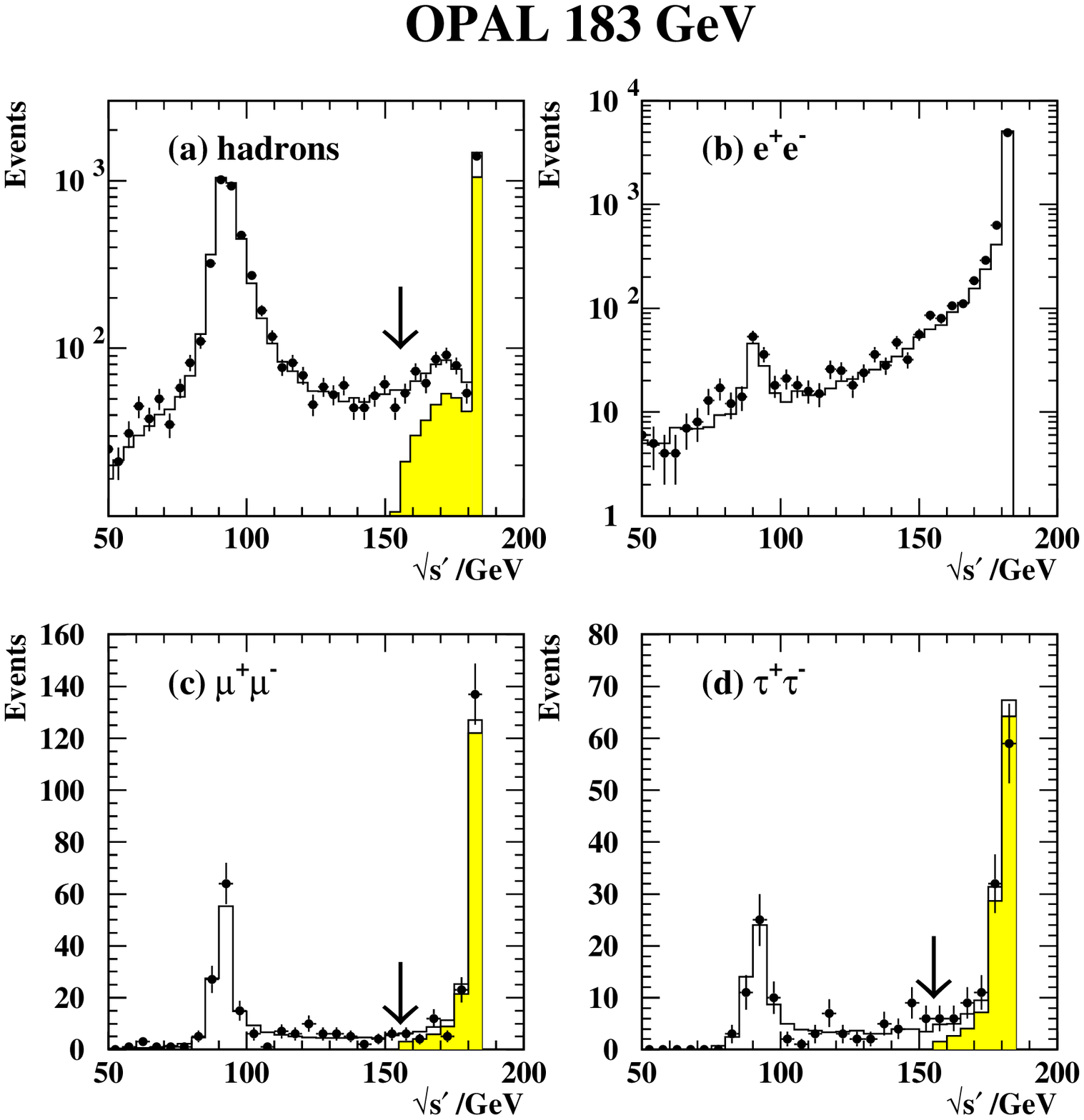}
  \caption
{
 The distributions of reconstructed $\protect\sqrt{s'}$ 
 for (a) hadronic events, (b) electron pair events with $\absctep < 0.9$, 
 $\absctem < 0.9$ and $\thacol < 170\degree$,
 (c) muon pair and (d) tau pair events at 182.69~GeV. In each case, the points 
 show the data and the histogram the Monte Carlo prediction, normalized
 to the integrated luminosity of the data, with the 
 contribution from events with true $s'/s > 0.7225$ shaded in (a), (c) and (d).
 The arrows in (a), (c) and (d) show the position of the cut used to
 select `non-radiative' events. 
}
\label{fig:sp}
\end{figure}

\begin{figure}
  \epsfxsize=\textwidth
  \epsfbox[0 0 567 680]{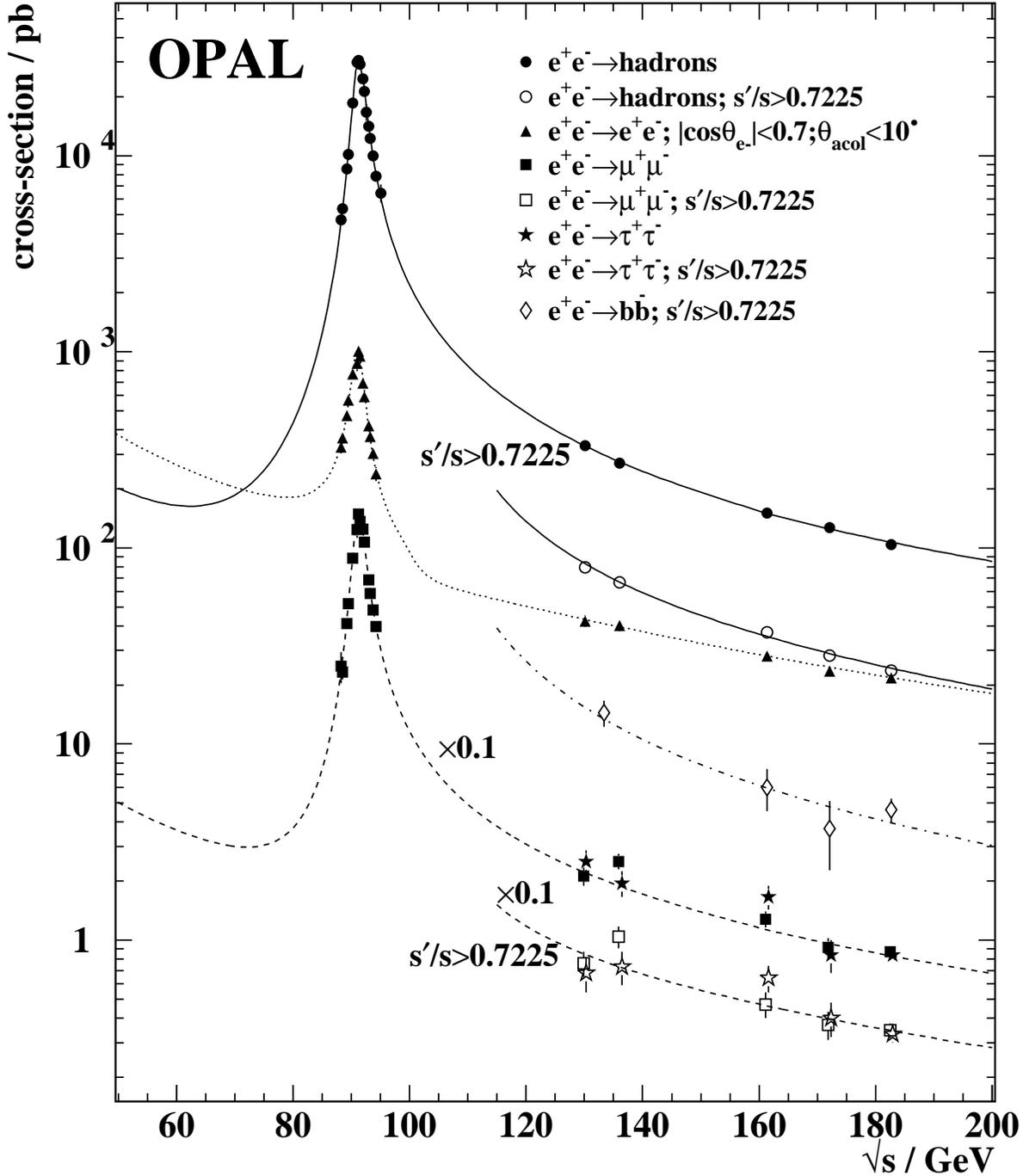}
  \caption
{
    Measured total cross-sections ($s'/s>0.01$) for different final
    states at lower energies~\cite{bib:OPAL-LS90,bib:OPAL-LS91,bib:OPAL-LS92,
    bib:OPAL-SM172}, and this analysis. Cross-section measurements for 
    hadrons, \bbbar, \mumu\ and \tautau\ for $s'/s>0.7225$ from this analysis 
    and from~\cite{bib:OPAL-SM172} are also shown; the latter have been
    corrected from $s'/s > 0.8$ to $s'/s > 0.7225$ by adding the prediction
    of ZFITTER for this difference before plotting. The cross-sections for 
    \Pgmp\Pgmm\ and \Pgtp\Pgtm\ production have been reduced by a factor of 
    ten for clarity. The curves show the predictions of ZFITTER for hadronic 
    (solid), \bbbar\ (dot-dashed), \Pgmp\Pgmm\ and \Pgtp\Pgtm\ (dashed) 
    final states and that of ALIBABA for the \Pep\Pem\ final state (dotted). 
}
\label{fig:xsec}
\end{figure}

In~\cite{bib:OPAL-SM172} non-radiative events were selected by demanding 
$s'/s > 0.8$; here we select non-radiative samples by demanding 
$s'/s > 0.7225$. Distributions of $\sqrt{s'}$ for each channel, determined 
using kinematic fits for hadrons and track angles for the lepton pairs as
in~\cite{bib:OPAL-SM172}, are shown in Fig.~\ref{fig:sp}.
Efficiencies, backgrounds and feedthrough of events from lower $s'$ into the 
non-radiative samples were calculated from Monte Carlo simulation, and
are given in Table~\ref{tab:eff}. Efficiencies determined from
two-fermion Monte Carlo events have been corrected for the effect of the
four-fermion contribution as described in~\cite{bib:OPAL-SM172}.
The numbers of selected events and the measured cross-sections are presented 
in Table~\ref{tab:xsec_183}, and the cross-sections shown in
Fig.~\ref{fig:xsec}. As well as cross-sections for \qqbar\ events, 
we also present a fully inclusive hadronic cross-section 
$\sigma(\qqbar\mathrm{X})$. This uses the same event selection as is
used for \qqbar\ events but W-pairs are not rejected. The cross-section
therefore includes W-pair production with at least one W decaying hadronically.
All cross-sections except those for \epem\ have been corrected for the 
contribution of interference between initial- and final-state radiation as 
described in~\cite{bib:OPAL-SM172}. The corrections are shown in 
Table~\ref{tab:ifsr}.

\begin{table}[htbp]
\centering
\begin{tabular}{|ll|c|c|c|}
\hline
\hline
\multicolumn{5}{|c|}{\bf Efficiencies and backgrounds at $\sqrt{s}$ = 183~GeV}
\\
\hline
Channel  & &Efficiency (\%)  &Background (pb) &Feedthrough (pb) \\ 
\hline
\qqbar X & &91.0$\pm$1.2 &4.5$\pm$1.0 &-- \\
\hline
\qqbar &$s'/s>0.01$ &88.6$\pm$1.3 &6.4$\pm$1.0 &-- \\
       &$s'/s>0.7225$ & 89.9$\pm$0.9 &2.09$\pm$0.09 &1.2$\pm$0.2 \\ 
\hline
\Pep\Pem &$\absct<0.9$, $\thacol<170\degree$  &98.1$\pm$1.1 
                                               &1.6$\pm$0.1   &-- \\
         &$\absctem<0.7$, $\thacol<10\degree$ &99.1$\pm$0.6 
                                               &0.26$\pm$0.05 &-- \\
         &$\absct<0.96$, $\thacol<10\degree$  &98.9$\pm$1.0 
                                               &10.9$\pm$1.1  &-- \\
\hline
\Pgmp\Pgmm &$s'/s>0.01$ &74.2$\pm$0.8 &0.31$\pm$0.08 &-- \\
           &$s'/s>0.7225$ &88.3$\pm$0.9 &0.06$\pm$0.02 &0.065$\pm$0.002 \\
\hline
\Pgtp\Pgtm &$s'/s>0.01$ &40.4$\pm$0.9 &0.61$\pm$0.11 &-- \\
           &$s'/s>0.7225$ &58.9$\pm$1.3 &0.22$\pm$0.04 &0.085$\pm$0.003 \\
\hline
\hline
\end{tabular}
\caption[]{
  Efficiency of selection cuts, background and feedthrough of events
  with lower $s'$ into the non-radiative samples for each channel at 
  183~GeV. The errors include Monte Carlo statistics and systematic 
  effects. In the case of electron pairs, the efficiencies are
  effective values including the efficiency of selection cuts for events
  within the acceptance region and the effect of acceptance corrections.  
  Values for \bbbar\ production are given in the text.
}
\label{tab:eff}
\end{table}

\begin{table}[htbp]
\centering
\begin{tabular}{|ll|c|r|l|c|}
\hline
\hline
\multicolumn{6}{|c|}{\bf Cross-sections at $\sqrt{s}$ = 183~GeV} \\
\hline
Channel   &       &$\int{\cal L}\mathrm{d}t$ (pb$^{-1}$) &  Events & 
    \multicolumn{1}{c|}{$\sigma$ (pb)} &  $\sigma^{\mathrm{SM}}$ (pb) \\
\hline
\qqbar X  & &56.9 & 6373  & 118.3$\pm$1.5$\pm$1.6   & 121.3   \\
\hline
\qqbar &$s'/s>0.01$   &56.9 & 5598  & 103.8$\pm$1.5$\pm$1.7   & 106.7 \\
       &$s'/s>0.7225$ &57.8 & 1408 & 23.7$\pm$0.7$\pm$0.4 & 24.3 \\
\hline
b$\overline{\rm b}$ &$s'/s>0.7225$ &55.5 & 
             348 -- 102 & 4.6 $\pm$0.6$\pm$0.3   & 3.96 \\
\hline
\Pep\Pem &$\absct<0.9$, $\thacol<170\degree$  &57.8 & 6980 &
                                              121.4$\pm$1.5$\pm$1.5 & 120.0 \\
         &$\absctem<0.7$, $\thacol<10\degree$ &     & 1260 &
                                               21.7$\pm$0.6$\pm$0.2 & 21.8 \\
         &$\absct<0.96$, $\thacol<10\degree$  &     & 19641 &
                                                333$\pm$3$\pm$4 & 333 \\
\hline
\Pgmp\Pgmm &$s'/s>0.01$   &53.9 &366 &8.70$\pm$0.46$\pm$0.19  & 8.31 \\
           &$s'/s>0.7225$ &     &174 &3.46$\pm$0.26$\pm$0.12  & 3.45 \\
\hline
\Pgtp\Pgtm &$s'/s>0.01$   &53.9 &216 &8.38$\pm$0.57$\pm$0.34  & 8.30 \\
           &$s'/s>0.7225$ &     &123 &3.31$\pm$0.30$\pm$0.11  & 3.45 \\
\hline
\hline
\end{tabular}
\caption[]{
  Integrated luminosity used in the analysis, numbers of selected events 
  and measured cross-sections at $\sqrt{s}$=182.69~GeV. In the \bbbar\ 
  case the numbers of forward and backward tags are given.
  For the cross-sections, the first error shown is statistical, the second
  systematic.  As in~\cite{bib:OPAL-SM172}, the cross-sections
  for hadrons, b$\overline{\rm b}$, \Pgmp\Pgmm\ and \Pgtp\Pgtm\ 
  are defined to cover phase-space up to the limit imposed by the $s'/s$ cut, 
  with $\sqrt{s'}$ defined as the invariant mass
  of the outgoing two-fermion system {\em before} final-state photon
  radiation. The contribution of interference between initial-
  and final-state radiation has been removed.
  The last column shows the Standard Model cross-section predictions from
  ZFITTER~\cite{bib:zfitter} (hadrons, \bbbar, \Pgmp\Pgmm, \Pgtp\Pgtm)
  and  ALIBABA~\cite{bib:alibaba} (\Pep\Pem).
}
\label{tab:xsec_183}
\end{table}

\begin{table}[htbp]
\centering
\begin{tabular}{|l|l|l|l|}
\hline
\hline
\multicolumn{4}{|c|}{\bf Interference Corrections } \\
\hline
\boldmath $s'/s>0.01$  & 130.12~GeV       & 136.08~GeV       & 182.69~GeV  \\
\hline
$\Delta\sigma/\sigma_{\rm SM}$(had) (\%)  
    & $+0.10\pm0.00\pm0.10$ & $+0.10\pm0.00\pm0.10$ & $+0.14\pm0.00\pm0.14$ \\
$\Delta\sigma/\sigma_{\rm SM}(\mu\mu)$ (\%)
             & $-0.35\pm0.05$     & $-0.33\pm0.03$     & $-0.43\pm0.04$\\
$\Delta\sigma/\sigma_{\rm SM}(\tau\tau)$ (\%)
             & $-0.44\pm0.09$     & $-0.40\pm0.07$     & $-0.46\pm0.06$\\
$\Delta A_{\rm FB}(\mu\mu)$
             & $-0.0040\pm0.0006$ & $-0.0029\pm0.0003$ & $-0.0052\pm0.0007$\\
$\Delta A_{\rm FB}(\tau\tau)$
             & $-0.0055\pm0.0014$ & $-0.0056\pm0.0013$ & $-0.0055\pm0.0009$\\
$\Delta A_{\rm FB}$(combined)
             & $-0.0046\pm0.0008$ & $-0.0036\pm0.0006$ & $-0.0055\pm0.0008$\\
\hline
\boldmath $s'/s>0.7225$  & 130.12~GeV     & 136.08~GeV       & 182.69~GeV  \\
\hline
$\Delta\sigma/\sigma_{\rm SM}$(had) (\%)  
             & $+0.8\pm0.2\pm0.4$ & $+0.9\pm0.3\pm0.4$ & $+1.2\pm0.3\pm0.6$ \\
$\Delta\sigma/\sigma_{\rm SM}(\mu\mu)$ (\%)
             & $-1.4\pm0.4$       & $-1.4\pm0.4$       & $-1.4\pm0.4$\\
$\Delta\sigma/\sigma_{\rm SM}(\tau\tau)$ (\%)
             & $-1.2\pm0.4$       & $-1.1\pm0.3$       & $-1.2\pm0.3$\\
$\Delta A_{\rm FB}(\mu\mu)$
             & $-0.015\pm0.004$   & $-0.007\pm0.002$   & $-0.015\pm0.004$\\
$\Delta A_{\rm FB}(\tau\tau)$
             & $-0.011\pm0.004$   & $-0.008\pm0.002$   & $-0.012\pm0.004$\\
$\Delta A_{\rm FB}$(combined)
             & $-0.014\pm0.004$   & $-0.007\pm0.002$   & $-0.014\pm0.004$\\  
\hline
\boldmath $s'/s>0.7225$ & \multicolumn{2}{c|}{133.29~GeV} & 182.69~GeV  \\
\hline
$\Delta\sigma/\sigma_{\rm SM}(\bbbar)$ (\%)  
             & \multicolumn{2}{c|}{ $-1.0\pm0.3\pm0.4$} & $-1.4\pm0.4\pm0.6$ \\
$\Delta R_{\rm b}/R_{\rm b,SM}$ (\%)  
             & \multicolumn{2}{c|}{ $-1.7\pm0.5\pm0.6$} & $-2.3\pm0.6\pm1.1$ \\
\hline
\hline
\end{tabular}
\caption[]{Corrections $\Delta\sigma$ and $\Delta A_{\rm FB}$
 which have been applied to the measured cross-sections and 
 asymmetries in order to remove the contribution from interference
 between initial- and final-state radiation. Cross-section corrections are
 expressed as a fraction of the expected Standard Model cross-section,
 while asymmetry corrections are given as absolute numbers, and
 depend on the observed asymmetry. The first error reflects the uncertainty 
 from  modelling the selection efficiency for the interference cross-section, 
 and is very small for hadrons because the efficiency is large and
 depends only weakly on \ct. The second error is our estimate of 
 possible additional QCD corrections for the hadrons~\cite{bib:OPAL-SM172}.
}
\label{tab:ifsr}
\end{table}

Systematic errors on these measurements are generally similar to
those at 172~GeV. In the case of inclusive hadrons they are
dominated by uncertainties in the selection efficiency (1.2\%) and
background from two-photon events (1.0\%), whereas the main uncertainties
for non-radiative hadronic events arise from their separation from 
radiative events (1.1\%) and knowledge of the efficiency (1.1\%).
As a cross-check, we have calculated the hadron cross-sections
without rejecting W-pair events, by subtracting their expected contribution
instead. The measured values of 103.9$\pm$1.6$\pm$1.6~pb ($s'/s > 0.01$)
and 23.7$\pm$0.8$\pm$0.5~pb ($s'/s > 0.7225$), after correction for 
interference between initial- and final-state radiation, are in good agreement 
with the values in Table~\ref{tab:xsec_183}. The main systematic
errors on the electron pair cross-sections arise from uncertainties
in the matching of central detector tracks to electromagnetic calorimeter
clusters, and knowledge of the acceptance correction and modelling
of the edge of the acceptance. Those on the muon and tau pair
cross-sections arise mainly from uncertainties in background and efficiency, 
and are substantially smaller than the statistical errors.

\begin{figure}
\epsfxsize=\textwidth
\epsfbox[0 0 567 680]{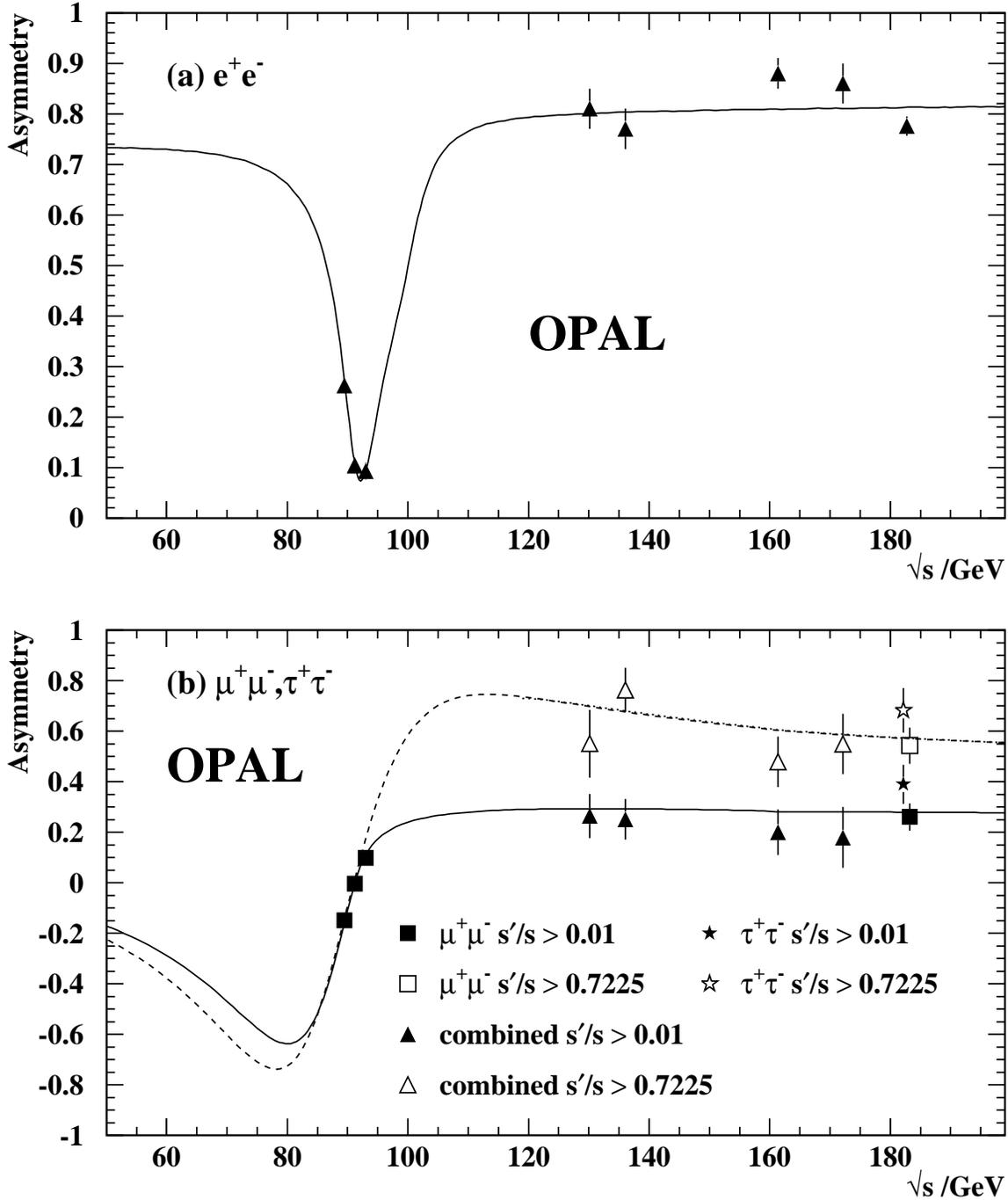}
\caption
{ (a) Measured forward-backward asymmetry for electron pairs with 
  $|\cos\theta_{\Pem}|<0.7$ and $\thacol<10\degree$, as a function
  of $\protect\roots$. The curve shows the prediction of ALIBABA.
  (b) Measured asymmetries for all ($s'/s>0.01$) and non-radiative
  ($s'/s>0.7225$) samples as functions of $\protect\roots$
  for \Pgmp\Pgmm\ and \Pgtp\Pgtm\ events.
  The curves show ZFITTER predictions for $s'/s>0.01$ (solid) and
  $s'/s>0.7225$ (dotted), as well as the Born-level expectation
  without QED radiative effects (dashed). The expectation for $s'/s>0.7225$ 
  lies very close to the Born curve, such that it appears 
  indistinguishable on this plot. 
}
\label{fig:afb}
\end{figure}

\begin{figure}
\epsfxsize=\textwidth
\epsfbox[0 0 567 567]{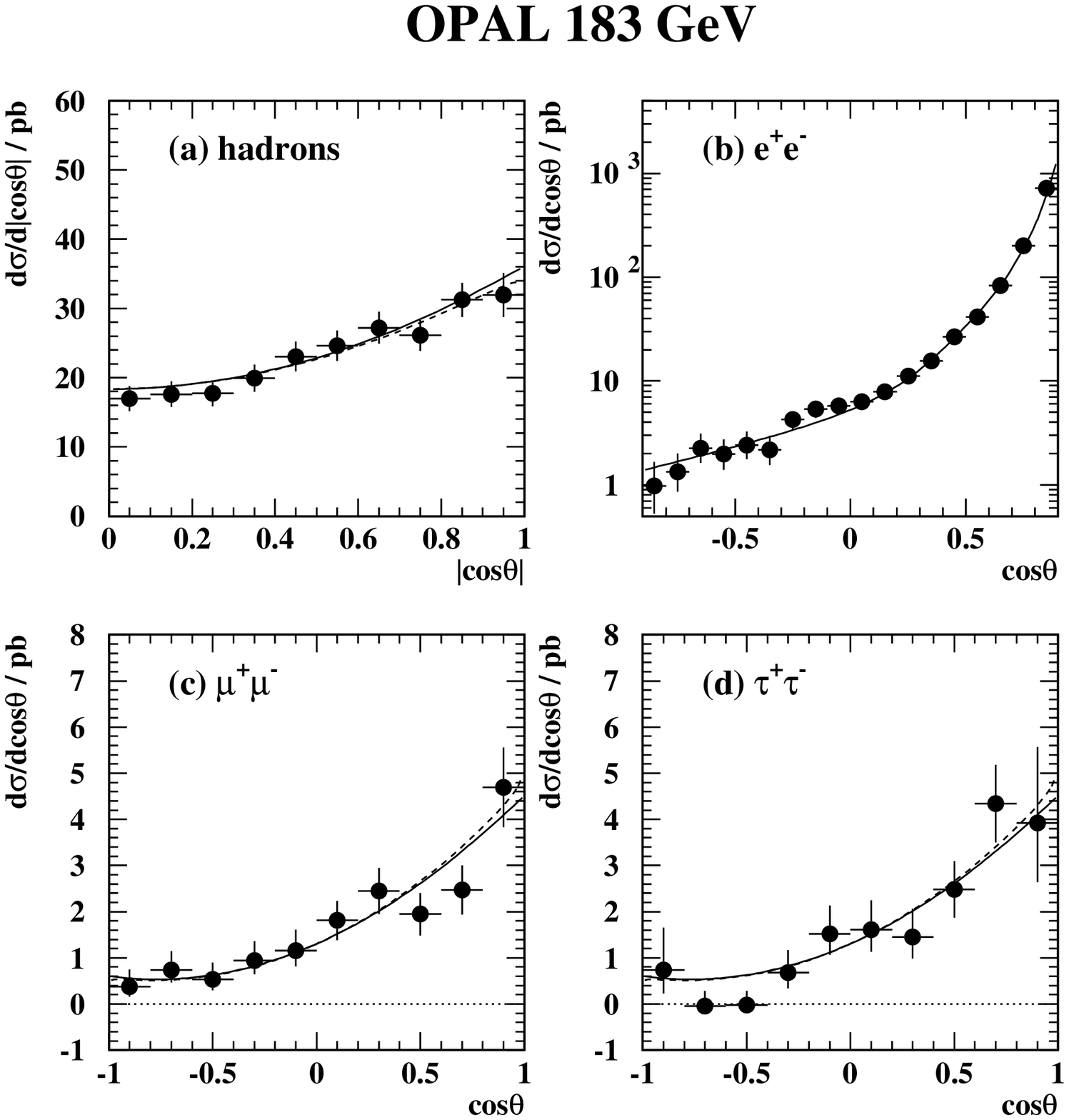}
\caption
{ Angular distributions for (a) hadronic events with $s'/s > 0.7225$,
 (b) \epem\ events with $\thacol < 10\degree$, (c) \mumu\ events with
 $s'/s > 0.7225$ and (d) \tautau\ events with $s'/s > 0.7225$.
 The points show the 183~GeV data, corrected to no interference between
 initial- and final-state radiation in (a), (c) and (d). The curve
 in (b) shows the prediction of ALIBABA, while the curves in (a),
 (c) and (d) show the predictions of ZFITTER with no interference
 between initial- and final-state radiation (solid) and with interference
 (dashed).
}
\label{fig:angdis}
\end{figure}

Measurements of the forward-backward asymmetry for lepton pairs are
given in Table~\ref{tab:afb_183} and compared with lower energy 
measurements in Fig.~\ref{fig:afb}. As before~\cite{bib:OPAL-SM172}, only 
events where the charge
can be reliably determined are used for the asymmetry measurements. The
final values for muon and tau pairs are obtained by averaging the results 
measured using the negative particle with those obtained using the positive 
particle to reduce systematic effects. Muon and tau asymmetries are corrected 
to the full angular range by applying a multiplicative correction obtained 
from ZFITTER to the asymmetry measured within the acceptance of the selection
cuts.  As at lower energies, the dominant
errors on the asymmetry measurements are statistical, with systematic 
effects from the correction for interference between initial- and final-state 
radiation, charge misassignment and $\theta$ measurement amounting to 0.01
or less in all cases. The corrected angular distributions for the
lepton pairs are given in Table~\ref{tab:angdis}. The angular distribution
of the primary quark in non-radiative hadronic events is given in
Table~\ref{tab:mh_angdis}. The angular distributions are plotted in
Fig.~\ref{fig:angdis}.

\begin{table}[htbp]
\centering
\begin{tabular}{|ll|r|r|c|c|}
\hline
\hline
\multicolumn{6}{|c|}{\bf Asymmetries at $\sqrt{s}$ = 183~GeV} \\
\hline
&&$N_{\mathrm{F}}$ &$N_{\mathrm{B}}$ &$\AFB$ &$\AFBSM$ \\ 
\hline
\epem    &$\absctem < 0.7$ & 1088      & 140     &0.776$\pm$0.019 &0.813 \\
         &and $\thacol < 10^\circ$ & & & &                \\ 
\hline
\mumu    &$s'/s > 0.01$    &229        &122     &0.26$\pm$0.05  &0.28 \\
         &$s'/s > 0.7225$  &127.5      &38.5    &0.54$\pm$0.07  &0.57 \\
\hline
\tautau  &$s'/s > 0.01$    &149        &58      &0.39$\pm$0.08  &0.28 \\
         &$s'/s > 0.7225$  &97         &23      &0.68$\pm$0.09  &0.57 \\
\hline
Combined &$s'/s > 0.01$    &           &        &0.31$\pm$0.04  &0.28 \\
\mumu\ and \tautau &$s'/s > 0.7225$  & &        &0.60$\pm$0.05  &0.57 \\
\hline
\hline
\end{tabular}
\caption[]{The numbers of forward ($N_{\mathrm{F}}$) and backward
  ($N_{\mathrm{B}}$) events and measured asymmetry values at 182.69~GeV. The
  measured asymmetry values include corrections for background and efficiency,
  and in the case of muons and taus are corrected to the full solid angle.
  The errors shown are the combined statistical and systematic errors.
  The asymmetries for
  \Pgmp\Pgmm, \Pgtp\Pgtm\ and for the combined \mumu\ and \tautau\ are
  shown after the correction for interference between
  initial- and final-state radiation.
  The final column shows the Standard Model predictions of ALIBABA for
  \epem\ and ZFITTER for the other final states.
}
\label{tab:afb_183}
\end{table}

\begin{table}[p]
\begin{center}
\small
\begin{tabular} {|c|r@{$\pm$}l|r@{$\pm$}l|r@{$\pm$}l|}
\hline
\hline
\multicolumn{7}{|c|}{\boldmath $\epem$} \\
\hline
$[\ct_{\rm min},\ct_{\rm max}]$  &\multicolumn{6}{c|}{$\dsdcc$ (pb)} \\
\hline
                          &\multicolumn{2}{c|}{130.12~GeV}
                          &\multicolumn{2}{c|}{136.08~GeV}
                          &\multicolumn{2}{c|}{182.69~GeV} \\
\hline
$[-0.9,-0.7]$ &  4&$^{3}_{2}$  &  5&$^{3}_{2}$  &1.2&$^{0.4}_{0.3}$ \\
$[-0.7,-0.5]$ &  4&$^{3}_{2}$  &  4&$^{3}_{2}$  &2.1&0.4 \\
$[-0.5,-0.3]$ &  6&$^{3}_{2}$  &  8&$^{4}_{3}$  &2.3&0.5 \\
$[-0.3,-0.1]$ &  6&$^{4}_{2}$  &  9&$^{4}_{3}$  &4.8&0.7 \\
$[-0.1,\;\;\;0.1]$ & 13&$^{5}_{4}$  &  8&$^{4}_{3}$  &6.1&0.7 \\
$[\;\;\;0.1,\;\;\;0.3]$ & 23&5           & 18&4  & 9.5&0.9 \\
$[\;\;\;0.3,\;\;\;0.5]$ & 45&7           & 35&6  &21.1&1.4 \\
$[\;\;\;0.5,\;\;\;0.7]$ &113&11          &122&10 & 62 &2   \\
$[\;\;\;0.7,\;\;\;0.9]$ &839&30          &725&27 &458 &8    \\
\hline
\hline
\multicolumn{7}{|c|}{\boldmath $\mumu$} \\
\hline
$[\ct_{\rm min},\ct_{\rm max}]$  &\multicolumn{6}{c|}{$\dsdcc$ (pb)} \\
\hline
                          &\multicolumn{2}{c|}{130.12~GeV}
                          &\multicolumn{2}{c|}{136.08~GeV}
                          &\multicolumn{2}{c|}{182.69~GeV} \\
\hline
$[-1.0,-0.8]$ &$0$&$^{3}_{1}$  &$-1$&$^{1}_{0}$  &$0.4$&$^{0.4}_{0.2}$ \\
$[-0.8,-0.6]$ &$3$&$^{3}_{2}$  &$2$&$^{3}_{1}$   &$0.7$&$^{0.4}_{0.3}$ \\
$[-0.6,-0.4]$ &$0$&$^{2}_{1}$  &$0$&$^{2}_{1}$   &$0.5$&$^{0.4}_{0.2}$ \\
$[-0.4,-0.2]$ &$6$&$^{4}_{3}$  &$5$&$^{3}_{2}$   &$0.9$&$^{0.4}_{0.3}$ \\
$[-0.2,\;\;\;0.0]$  &$3$&$^{3}_{2}$  &$1$&$^{2}_{1}$  &$1.2$&$^{0.5}_{0.3}$ \\
$[\;\;\;0.0,\;\;\;0.2]$ &$2$&$^{3}_{2}$  &$3$&$^{3}_{2}$  &$1.8$&$0.4$ \\
$[\;\;\;0.2,\;\;\;0.4]$ &$6$&$^{4}_{3}$  &$6$&$^{3}_{2}$  &$2.4$&$0.5$ \\
$[\;\;\;0.4,\;\;\;0.6]$ &$5$&$^{4}_{2}$  &$10$&$^{4}_{3}$ &$1.9$&$0.5$ \\
$[\;\;\;0.6,\;\;\;0.8]$ &$5$&$^{4}_{3}$  &$ 9$&$^{4}_{3}$ &$2.5$&$0.5$ \\
$[\;\;\;0.8,\;\;\;1.0]$ &$9$&$^{5}_{4}$  &$18$&$^{7}_{5}$ &$4.7$&$0.9$ \\
\hline
\hline
\multicolumn{7}{|c|}{\boldmath $\tautau$} \\
\hline
$[\ct_{\rm min},\ct_{\rm max}]$  &\multicolumn{6}{c|}{$\dsdcc$ (pb)} \\
\hline
                          &\multicolumn{2}{c|}{130.12~GeV}
                          &\multicolumn{2}{c|}{136.08~GeV}
                          &\multicolumn{2}{c|}{182.69~GeV} \\
\hline
$[-1.0,-0.8]$ &$-1$&$^{10}_{0}$ &$-1$&$^{9}_{0}$  &$0.7$&$^{0.9}_{0.5}$ \\
$[-0.8,-0.6]$ &$0$&$^{1}_{0}$   &$0$&$^{1}_{0}$   &$0.0$&$^{0.3}_{0.1}$ \\
$[-0.6,-0.4]$ &$-1$&$^{2}_{0}$  &$2$&$^{3}_{2}$   &$0.0$&$^{0.3}_{0.1}$ \\
$[-0.4,-0.2]$ &$2$&$^{4}_{2}$   &$0$&$^{1}_{0}$   &$0.7$&$^{0.5}_{0.3}$ \\
$[-0.2,\;\;\;0.0]$ &$3$&$^{4}_{2}$       &$1$&$^{3}_{1}$   
                                          &$1.5$&$^{0.6}_{0.5}$ \\
$[\;\;\;0.0,\;\;\;0.2]$ &$2$&$^{4}_{2}$       &$4$&$^{3}_{2}$   
                                          &$1.6$&$^{0.6}_{0.5}$ \\
$[\;\;\;0.2,\;\;\;0.4]$ &$1$&$^{4}_{2}$       &$4$&$^{4}_{2}$   
                                          &$1.5$&$^{0.6}_{0.5}$ \\
$[\;\;\;0.4,\;\;\;0.6]$ &$ 7$&$^{5}_{3}$      &$7 $&$^{4}_{3}$  &$2.5$&$0.6$ \\
$[\;\;\;0.6,\;\;\;0.8]$ &$7$&$^{5}_{3}$       &$8$&$^{5}_{3}$   &$4.3$&$0.8$ \\
$[\;\;\;0.8,\;\;\;1.0]$ &$14$&$^{19}_{11}$    &$7$&$^{15}_{7}$ 
                                          &$3.9$&$^{1.6}_{1.3}$ \\
\hline
\hline
\end{tabular}     
\caption{
Differential cross-sections for lepton pair production. The values for
$\epem$ are for $\thacol < 10\degree$; those for $\mumu$ and $\tautau$
are for $s'/s > 0.7225$ and are corrected to no interference between 
initial- and final-state radiation. The values for 130 and 136~GeV are
from the combined 1995 and 1997 data. Negative values arise when fewer
events are observed than expected from background. Errors include statistical 
and systematic effects combined, with the former dominant.}
\label{tab:angdis}
\end{center}  
\end{table}

\begin{table}[htbp]
\begin{center}
\begin{tabular} {|c|c|c|c|}
\hline
\hline
\multicolumn{4}{|c|}{\bf \boldmath Hadrons} \\
\hline
$\absct$   &\multicolumn{3}{c|}{$\dsdabscc$ (pb)} \\
\hline
           &130.12~GeV &136.08~GeV &182.69~GeV  \\
\hline
$ [0.0,0.1]$ & 70$\pm$12 & 48$\pm$9  &17.0$\pm$1.8 \\
$ [0.1,0.2]$ & 52$\pm$9  & 64$\pm$10 &17.6$\pm$1.8 \\
$ [0.2,0.3]$ & 70$\pm$11 & 56$\pm$10 &17.7$\pm$1.9 \\
$ [0.3,0.4]$ & 70$\pm$11 & 63$\pm$10 &19.9$\pm$2.0 \\
$ [0.4,0.5]$ & 64$\pm$10 & 44$\pm$9  &23.1$\pm$2.1 \\
$ [0.5,0.6]$ & 79$\pm$12 & 39$\pm$8  &24.6$\pm$2.2 \\
$ [0.6,0.7]$ & 81$\pm$12 & 78$\pm$11 &27.2$\pm$2.3 \\
$ [0.7,0.8]$ & 94$\pm$13 & 81$\pm$11 &26.1$\pm$2.2 \\
$ [0.8,0.9]$ & 85$\pm$12 & 82$\pm$11 &31.2$\pm$2.4 \\
$ [0.9,1.0]$ &160$\pm$23 &123$\pm$19 &32.0$\pm$3.2 \\
\hline
\hline
\end{tabular}     
\caption{
Differential cross-sections for $\qqbar$ production, for $s'/s > 0.7225$.  
The values are corrected to no interference between initial- and final-state 
radiation as in~\cite{bib:OPAL-SM172}. Errors include statistical and 
systematic effects combined, with the former dominant.}
\label{tab:mh_angdis}
\end{center}  
\end{table}

We have measured \Rb, the ratio of the cross-section for \bbbar\
production to the hadronic cross-section, for $s'/s > 0.7225$.
As in~\cite{bib:OPAL-SM172}, \bbbar\ events were tagged by
reconstructing secondary vertices in the plane transverse to the
beam direction, and a `folded tag' technique was used. In this method
\Rb\ is determined from the difference between the number of events
with $L/\sigma_L > 3$ (forward tags) and the number with 
$L/\sigma_L < -3$ (backward tags), where $L$ is the signed decay length 
measured from the primary vertex to the secondary vertex and $\sigma_L$ 
is the error on this length. The net efficiencies, i.e.\ differences between 
forward and
backward tagging efficiencies, determined from Monte Carlo, were
0.398$\pm$0.021, 0.095$\pm$0.004 and 0.0105$\pm$0.0024 for b, c and
light quarks respectively, similar to the values at 172~GeV. The systematic
errors on \Rb\ are again dominated by modelling of b and c fragmentation 
and decay, estimated using the prescription in~\cite{bib:rblep1},
and track parameter resolution.

This measurement used hadronic events with $s'/s > 0.7225$, selected
as for the cross-section measurement. W-pair events were rejected as
described above. A new background arises at 183~GeV from Z-pair production; 
the probability of a Z-pair event being tagged was estimated to be 
(33.5$\pm$1.0)\%, and the expected contribution from these was subtracted; 
this contribution is small, amounting to only 1.6\% of the tagged sample. 
After background subtraction the b purity of the tagged \qqbar\ sample was 
estimated to be 68\%.

We obtain a value of \Rb, after correcting for interference between
initial- and final-state radiation, of
\[
   \Rb (\roots = 182.69~\mathrm{GeV}) = 0.195\pm0.023\pm0.013,
\]
where the first error is statistical and the second systematic; the value
of the \bbbar\ cross-section derived from the measurement of the hadronic
cross-section and \Rb\ is given in Table~\ref{tab:xsec_183}.

\subsection{\bf Measurements at $\protect\sqrt{s}$ = 130--136~GeV} 
The analyses of the 130 and 136~GeV data were unchanged from those
described in~\cite{bib:OPAL-SM172}, with the exception of the modification
of the $s'$ cut used to define non-radiative events. The numbers of
selected events and measured cross-sections are given in 
Table~\ref{tab:xsec_130}, and the forward-backward asymmetry values
for the leptonic channels in Table~\ref{tab:afb_130}. The values are
generally in good agreement with those obtained from the data
collected in 1995 at similar centre-of-mass energies~\cite{bib:OPAL-SM172},
after correcting the latter to $s'/s > 0.7225$ in the non-radiative
cases. We have combined the measurements from the two sets of data;
when measurements of non-radiative hadrons, muons and taus were
combined the new $s'$ cut was applied to the 1995 data.
The combined results are also shown in Tables~\ref{tab:xsec_130}
and~\ref{tab:afb_130}. Corrected angular distributions for the
lepton pairs are given in Table~\ref{tab:angdis} and for the hadronic
events in Table~\ref{tab:mh_angdis}.

The measurement of \Rb\ for the 1997 data benefited, in comparison
with the 1995 measurement, from the longer silicon microvertex detector 
installed in 1996, which enabled the cut on the polar angle of the thrust 
axis to be increased from $\absct < 0.8$ to $\absct < 0.9$, as for
the measurements at 161--183~GeV. For the 1997 combined 130 and 136~GeV data 
we obtain a value of \Rb, after correction for initial-final state 
interference, of
\[
   \Rb (\roots = 133.38~\mathrm{GeV}) = 0.201\pm0.036\pm0.013,
\]
where the first error is statistical and the second systematic. Combining
with the 1995 measurement~\cite{bib:OPAL-SM172} and taking account of
correlated systematic errors, gives
\[
   \Rb (\roots = 133.29~\mathrm{GeV}) = 0.198\pm0.026\pm0.013.
\]
The corresponding \bbbar\ cross-sections, calculated from the hadronic
cross-sections and the \Rb\ measurements, are given in 
Table~\ref{tab:xsec_130}.
 
\begin{table}[htbp]
\centering
\begin{tabular}{|ll|c|c|c|c|}
\hline
\hline
\multicolumn{6}{|c|}{\bf Cross-sections at $\sqrt{s}$ = 130~GeV} \\
\hline
        &&\multicolumn{2}{c|}{1997}        &1995+1997  & \\
        &&\multicolumn{2}{c|}{130.00~GeV}  &130.12~GeV &130.12~GeV \\
        \cline{3-6}
Channel &&Events  &$\sigma$ (pb) &$\sigma$ (pb) &$\sigma^{\mathrm{SM}}$ (pb) \\
\hline
\qqbar &$s'/s>0.01$  & 807   &350$\pm$13$\pm$5 &332$\pm$8$\pm$5 &331 \\
       &$s'/s>0.7225$ & 241 
      &88.5$\pm$5.8$\pm$1.8  &79.7$\pm$3.8$\pm$1.6  &83.1 \\
\hline
\Pep\Pem &$\absct<0.9$, $\thacol<170\degree$   &609 
                         &235$\pm$10$\pm$4     &227$\pm$7$\pm$3 &238 \\
         &$\absctem<0.7$, $\thacol<10\degree$  & 113  
                         &43.2$\pm$4.1$\pm$0.6 &42.3$\pm$2.9$\pm$0.5 &43.2 \\
         &$\absct<0.96$, $\thacol<10\degree$   & 1718  
                         &647$\pm$16$\pm$10    &631$\pm$11$\pm$8 &647 \\
\hline
\Pgmp\Pgmm &$s'/s>0.01$   &40  
      &18.1$\pm$2.9$\pm$0.7  &21.2$\pm$2.2$\pm$0.5  &22.1 \\
           &$s'/s>0.7225$ &17  
      &5.5 $\pm$1.4$\pm$0.5  &7.6 $\pm$1.1$\pm$0.3  &8.5 \\
\hline
\Pgtp\Pgtm &$s'/s>0.01$   &24  
      &22.5$\pm$4.6$\pm$0.7  &25.1$\pm$3.4$\pm$0.8  &22.1 \\
           &$s'/s>0.7225$ &11  
      &6.4 $\pm$1.9$\pm$0.3  &6.8 $\pm$1.4$\pm$0.3  &8.5 \\
\hline
\hline
\multicolumn{6}{|c|}{\bf Cross-sections at $\sqrt{s}$ = 136~GeV} \\
\hline
        &&\multicolumn{2}{c|}{1997}        &1995+1997  & \\
        &&\multicolumn{2}{c|}{135.98~GeV}  &136.08~GeV &136.08~GeV \\
        \cline{3-6}
Channel &&Events  &$\sigma$ (pb)      &$\sigma$ (pb) &$\sigma^{\mathrm{SM}}$ \\
\hline
\qqbar &$s'/s>0.01$  & 907   &277$\pm$9$\pm$4 &271$\pm$7$\pm$4 &275 \\
       &$s'/s>0.7225$ & 228 
        &65.5$\pm$4.4$\pm$1.5  &66.7$\pm$3.3$\pm$1.3  &66.9 \\
\hline
\Pep\Pem &$\absct<0.9$, $\thacol<170\degree$   &704 
                         &209$\pm$8 $\pm$3     &204$\pm$6$\pm$3 &217 \\
\Pep\Pem &$\absctem<0.7$, $\thacol<10\degree$  & 144  
                         &42.4$\pm$3.6$\pm$0.6 &40.2$\pm$2.6$\pm$0.4 &39.6 \\
\Pep\Pem &$\absct<0.96$, $\thacol<10\degree$   & 2034  
                         &588$\pm$13$\pm$9     &585$\pm$10$\pm$7 &593 \\
\hline
\Pgmp\Pgmm &$s'/s>0.01$   &72  
      &25.0$\pm$3.0 $\pm$0.6  &25.2$\pm$2.2$\pm$0.5  &18.9 \\
           &$s'/s>0.7225$ &33  
      &9.5 $\pm$1.7 $\pm$0.4  &10.4$\pm$1.3$\pm$0.3  &7.3 \\
\hline
\Pgtp\Pgtm &$s'/s>0.01$   &23  
      &16.3$\pm$3.4 $\pm$0.6  &19.5$\pm$2.8$\pm$0.6  &18.9 \\
           &$s'/s>0.7225$ &15  
      &6.9 $\pm$1.8 $\pm$0.2  &7.3$\pm$1.4$\pm$0.3  &7.3 \\
\hline
\hline
\multicolumn{6}{|c|}{\bf Cross-sections at $\sqrt{s}$ = 133~GeV} \\
\hline
        &&\multicolumn{2}{c|}{1997}        &1995+1997  & \\
        &&\multicolumn{2}{c|}{133.38~GeV}  &133.29~GeV &133.29~GeV \\
        \cline{3-6}
Channel &&Events  &$\sigma$ (pb)      &$\sigma$ (pb) &$\sigma^{\mathrm{SM}}$ \\
\hline
b$\overline{\rm b}$ &$s'/s>0.7225$ & 
             126 -- 35 &15.2$\pm$2.8$\pm$1.0 &14.4$\pm$2.0$\pm$1.0 &13.5 \\
\hline
\hline
\end{tabular}
\caption[]{
  Numbers of selected events and measured cross-sections for the
  130 and 136~GeV data taken in 1997. In the \bbbar\ case the 130 and
  136~GeV data are combined, and numbers of forward and backward tags are 
  given. The cross-sections obtained by 
  combining these results with the results from data taken at these 
  energies in 1995~\cite{bib:OPAL-SM172} are also given, and compared with
  the Standard Model predictions from ALIBABA for electron pairs
  and ZFITTER for all other final states. Cross-sections are
  given after the correction for interference between initial- and 
  final-state radiation; values of these corrections are given in 
  Table~\ref{tab:ifsr}. The first error is statistical, the second 
  systematic.
}
\label{tab:xsec_130}
\end{table}

\begin{table}[htbp]
\centering
\begin{tabular}{|ll|r|r|c|c|c|}
\hline
\hline
\multicolumn{7}{|c|}{\bf Asymmetries at $\sqrt{s}$ = 130~GeV} \\
\hline
       &&\multicolumn{3}{c|}{1997}        &1995+1997  & \\
       &&\multicolumn{3}{c|}{130.00~GeV}  &130.12~GeV &130.12~GeV \\
       \cline{3-7}
&&$N_{\mathrm{F}}$ &$N_{\mathrm{B}}$ &$\AFB$ &$\AFB$\ &$\AFBSM$ \\ 
\hline
\epem    &$\absctem < 0.7$ &99         &9        
                           &0.84 $\pm$0.05  &0.81$\pm$0.04 &0.80 \\
         &and $\thacol < 10^\circ$ & & & & &               \\ 
\hline
\mumu    &$s'/s > 0.01$    &24.5       &15      
                           &0.24$\pm$0.16  &0.24$\pm$0.11  &0.29 \\
         &$s'/s > 0.7225$  &11.5       &5.5     
         &0.40$^{+0.30}_{-0.41}$       &0.40$^{+0.18}_{-0.22}$  &0.70 \\
\hline
\tautau  &$s'/s > 0.01$    &14         &8       
                           &0.25$\pm$0.23  &0.32$\pm$0.15  &0.29 \\
         &$s'/s > 0.7225$  &7          &3       
         &0.53$^{+0.35}_{-0.57}$       &0.80$^{+0.20}_{-0.31}$  &0.70 \\
\hline
Combined &$s'/s > 0.01$    &           &         
                           &0.24$\pm$0.14  &0.26$\pm$0.09  &0.29 \\
\mumu\ and \tautau &$s'/s > 0.7225$  & &        
                           &0.46$\pm$0.23  &0.55$\pm$0.13  &0.70 \\
\hline
\hline
\multicolumn{7}{|c|}{\bf Asymmetries at $\sqrt{s}$ = 136~GeV} \\
\hline
       &&\multicolumn{3}{c|}{1997}        &1995+1997  & \\
       &&\multicolumn{3}{c|}{135.98~GeV}  &136.08~GeV &136.08~GeV \\
       \cline{3-7}
&&$N_{\mathrm{F}}$ &$N_{\mathrm{B}}$ &$\AFB$ &$\AFB$\ &$\AFBSM$ \\
\hline
\epem    &$\absctem < 0.7$ &129        &15       
                           &0.79 $\pm$0.05  &0.77$\pm$0.04 &0.80 \\
         &and $\thacol < 10^\circ$ & & & & &               \\ 
\hline
\mumu    &$s'/s > 0.01$    &40.5       &29      
                           &0.14$\pm$0.12  &0.23$\pm$0.09  &0.29 \\
         &$s'/s > 0.7225$  &25.5       &7       
         &0.64$^{+0.16}_{-0.23}$       &0.71$^{+0.11}_{-0.14}$  &0.68 \\
\hline
\tautau  &$s'/s > 0.01$    &15         &7       
                           &0.23$\pm$0.25  &0.33$\pm$0.17  &0.29 \\
         &$s'/s > 0.7225$  &11.5       &3.5     
         &0.62$^{+0.26}_{-0.41}$       &0.86$^{+0.16}_{-0.26}$  &0.68 \\
\hline
Combined &$s'/s > 0.01$    &           &         
                           &0.15$\pm$0.11  &0.25$\pm$0.08  &0.29 \\
\mumu\ and \tautau &$s'/s > 0.7225$  & &        
                           &0.65$\pm$0.13  &0.76$\pm$0.09  &0.68 \\
\hline
\hline
\end{tabular}
\caption[]{The numbers of forward ($N_{\mathrm{F}}$) and backward
  ($N_{\mathrm{B}}$) events and measured asymmetry values for the 130 and
  136~GeV data taken in 1997. The values obtained by combining
  these results with the results from data taken at these energies in
  1995~\cite{bib:OPAL-SM172} are also given. Measured asymmetry values 
  include corrections for background and efficiency. In the case of muons and 
  taus they are corrected to the full solid angle, and are shown after the 
  correction for interference between initial- and final-state radiation.
  The errors shown are the combined statistical and systematic errors; in each
  case the systematic error is 0.01 or less. 
  The final column shows the Standard Model predictions of ALIBABA for
  \epem\ and ZFITTER for the other final states.
}
\label{tab:afb_130}
\end{table}

\section{Comparison with Standard Model Predictions}    \label{sec:sm}
The cross-section and asymmetry measurements at 183~GeV are compared
with the Standard Model predictions in Tables~\ref{tab:xsec_183} 
and~\ref{tab:afb_183} respectively. The Standard Model Predictions
are calculated using ALIBABA~\cite{bib:alibaba} for the \epem\ final
state and ZFITTER~\cite{bib:zfitter} for all other final states;
input parameters and flags used in ZFITTER are as in~\cite{bib:OPAL-SM172}.
The agreement is generally good, as is also seen in the case of the
130 and 136~GeV data in Tables~\ref{tab:xsec_130} and~\ref{tab:afb_130}.
Figure~\ref{fig:xsec} shows cross-sections, for both inclusive and
non-radiative events, as a function of $\roots$, while Fig.~\ref{fig:afb}
shows the measured asymmetry values. The angular distributions for
all channels at 183~GeV are compared with Standard Model predictions in
Fig.~\ref{fig:angdis}.

\begin{figure}
\epsfxsize=\textwidth
\epsfbox[0 0 567 567]{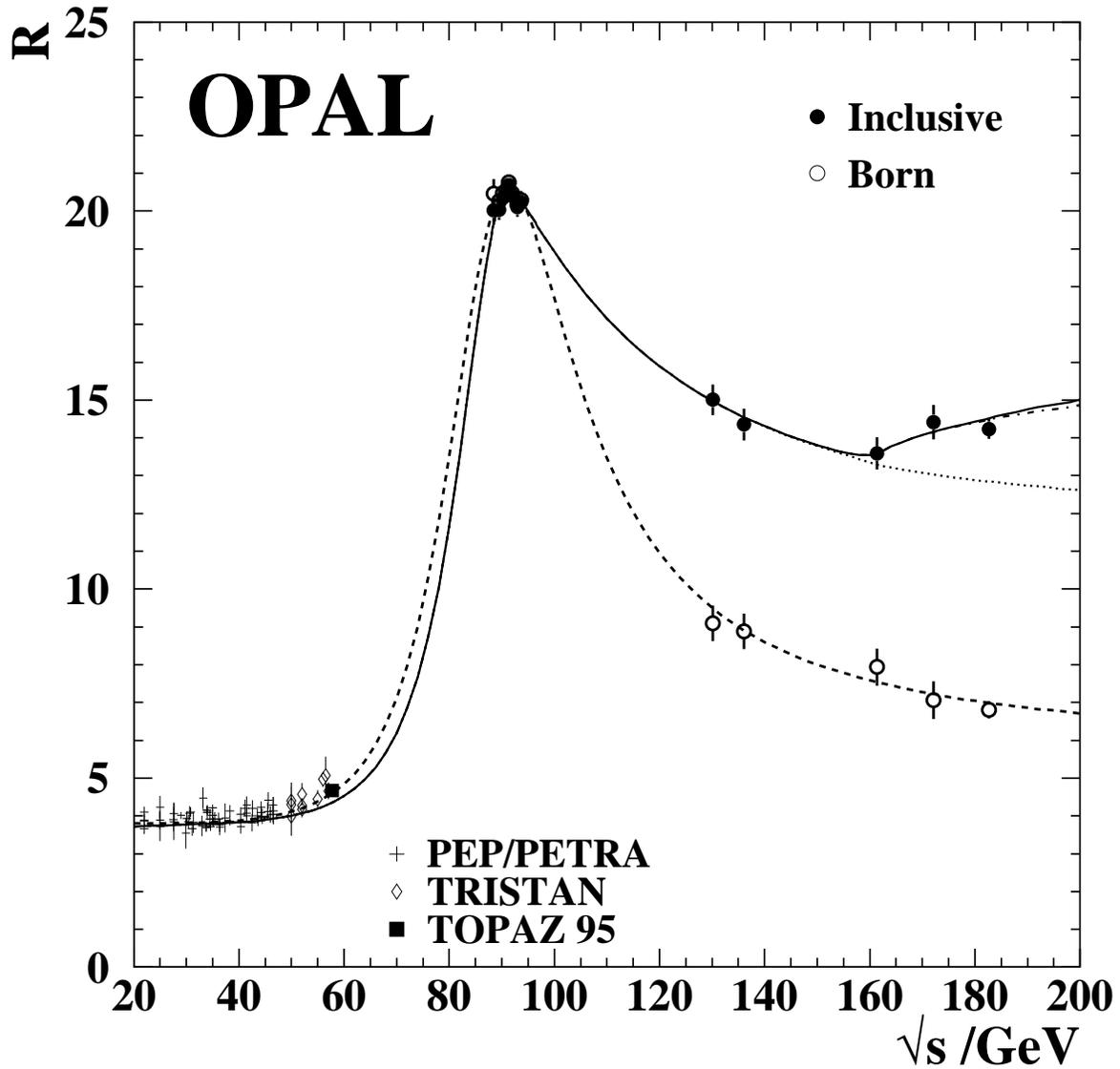}
\caption
{Ratio of measured hadronic cross-sections to theoretical muon
 pair cross-sections as a function of centre-of-mass energy. 
 Values are shown for the inclusive cross-section, $\sigma(\qqbar\mathrm{X})$
 and for the Born level cross-section. The dotted and dashed curves show 
 the predictions of ZFITTER for these cross-sections, 
 while the solid curve also includes the contributions from W-pairs 
 calculated using GENTLE~\cite{bib:gentle} and from Z-pairs calculated
 using FERMISV~\cite{bib:fermisv}. The dot-dashed curve is the total
 excluding the Z-pair contribution. Measurements at lower energies are from 
 references~\cite{bib:OPAL-SM172,bib:OPAL-LS90,bib:OPAL-LS91,bib:OPAL-LS92,
 bib:rdata}.
}
\label{fig:rplot}
\end{figure}

In Fig.~\ref{fig:rplot} we show the ratio of measured hadronic 
cross-sections to theoretical muon pair cross-sections as a function
of centre-of-mass energy for two cases. In the first case
the numerator of this ratio is the inclusive $\qqbar\mathrm{X}$ cross-section,
in the second case it is the non-radiative \qqbar\ cross-section
corrected to the Born level. In each case the denominator is the
corresponding muon pair cross-section calculated using ZFITTER. The inclusive
ratio clearly shows the effect of \WW\ production.

\subsection{\boldmath Energy Dependence of \alphaem} \label{sec:alphaem}
In~\cite{bib:OPAL-SM172} we used non-radiative cross-section and
asymmetry measurements to measure the electromagnetic coupling
constant \alphaem\ at LEP2 energies. We have repeated this fit
including the new measurements of hadronic, \mumu\ and \tautau\
cross-sections, \Rb, and the combined muon and tau asymmetry
values, for $s'/s > 0.7225$, presented here. As before, we form
the $\chi^2$ between the measured values and the Standard Model 
predictions calculated as a function of $\alphaem(\sqrt{s})$ using
ZFITTER, with all other ZFITTER input parameters fixed. Correlations
between measurements are fully taken into account.

We have performed separate fits to the 183~GeV measurements and the updated
130--136~GeV measurements; the 130 and 136~GeV measurements
are combined for this analysis. The results of these fits are given in
Table~\ref{tab:alphaem1}. The inclusion of the new data at 130--136~GeV
has not reduced the relative error on the measurement at 133~GeV 
because the measured cross-sections and asymmetries, particularly the 
hadronic cross-section, now lie closer to the Standard Model expectation. 
The sensitivity of the measurements to $\alphaem$ is nonlinear, resulting in 
smaller errors as the value moves away from the Standard Model expectation.
We have also performed a fit to data at all centre-of-mass energies, 
in which $\alphaem$ runs with energy with a slope corresponding to the 
fitted value. As input to the combined fit we use the measurements at 
183~GeV and the combined data at 130--136~GeV presented here, together with
measurements at 161 and 172~GeV from~\cite{bib:OPAL-SM172}.
The result of the combined fit is
\[
     1/\alphaem(169.34~\mathrm{GeV}) = 127.8^{+4.1}_{-3.6},
\]
where the value of \alphaem\ is quoted at the centre-of-mass energy
corresponding to the luminosity-weighted average of $1/s$. The 
errors on the fitted value of \alphaem\ arise from the errors on
the measurements; errors due to uncertainties in the ZFITTER input
parameters are negligible.

\begin{figure}
\epsfxsize=\textwidth
\epsfbox[0 0 567 567]{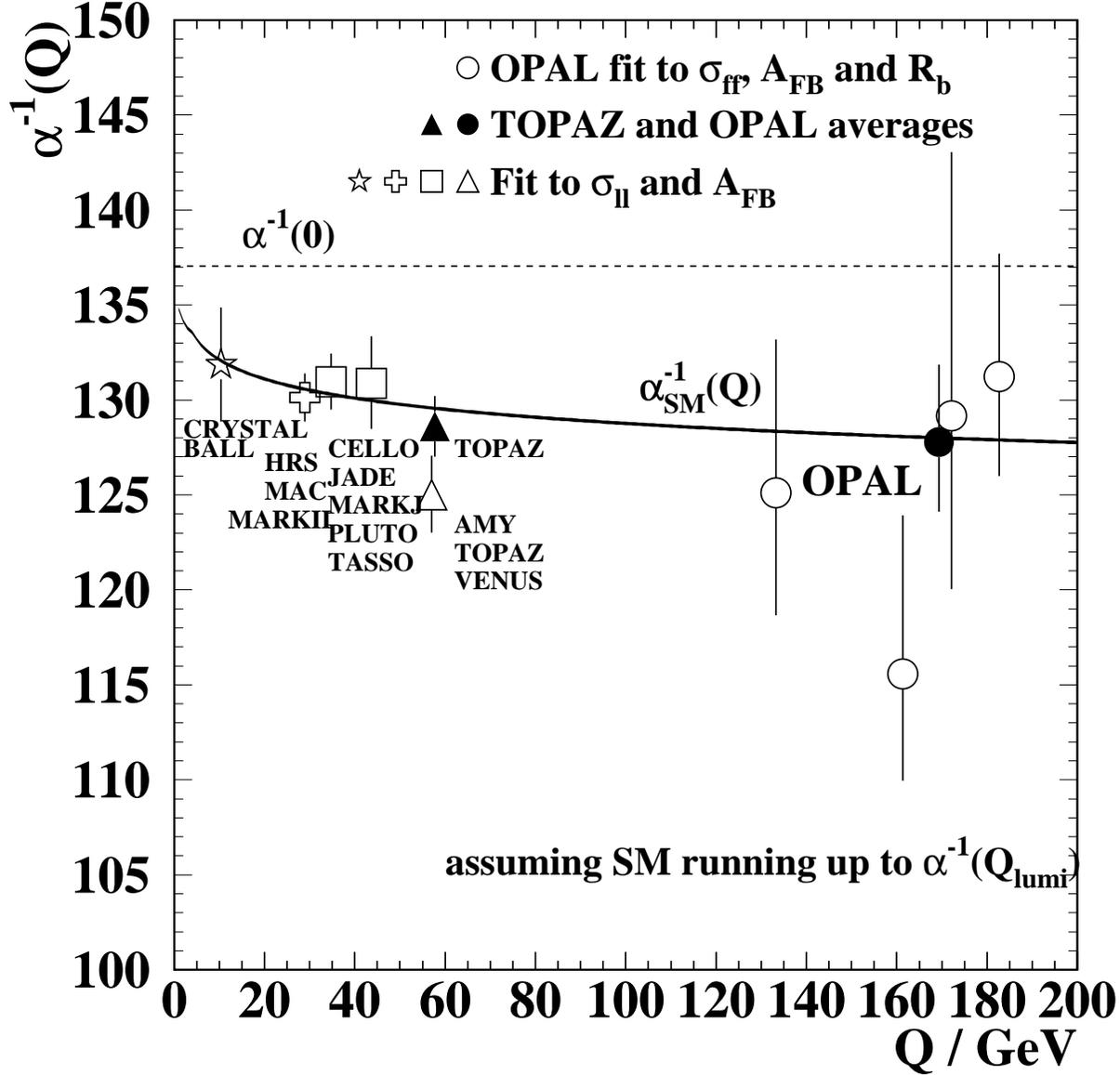}
\caption
{Fitted values of $1/\alphaem$ as a function of $Q$, which is
 $\protect \sqrt{s}$ for the OPAL fits. The open circles show the
 results of fits to OPAL data at each centre-of-mass energy, the closed
 circle the result of the combined fit in which \alphaem\ runs with a
 slope corresponding to its fitted value. The OPAL results at 161 and
 172~GeV are from~\cite{bib:OPAL-SM172}. Values obtained by the TOPAZ 
 experiment~\cite{bib:alrun} and from fits to measurements of leptonic
 cross-sections and asymmetries at the DORIS, PEP, PETRA and TRISTAN
 \epem\ storage rings~\cite{bib:MK_alphaem} are also shown. All
 measurements rely on assuming the Standard Model running of \alphaem\
 for $Q_{\mathrm{lumi}}$ below 5~GeV. 
 The solid line shows the Standard 
 Model expectation, with the thickness representing the uncertainty, 
 while the value of 1/$\alphaem(0)$ is shown by the dashed line. 
}
\label{fig:alphaem}
\end{figure}

The measured values of $\alphaem$ are shown in Fig.~\ref{fig:alphaem}.
They are consistent with the Standard Model expectation. The value
of $1/\alphaem$ obtained from the combined measurements is 2.3 standard 
deviations below the low energy limit of 
137.0359979$\pm$0.0000032~\cite{bib:CAGE}.

The fit described above uses measurements of cross-sections which 
depend on the measurement of luminosity, which itself assumes the
Standard Model running of $\alphaem$ from $(Q^2 = 0)$ to typically
$Q^2 = (3.5~\GeV)^2$, where $1/\alphaem\simeq 134$. Therefore it
measures the running of $\alphaem$ only from $Q_{\rm lumi}\simeq 3.5~\GeV$ 
onwards. To become independent of this assumption, as 
in~\cite{bib:OPAL-SM172} we have repeated the combined fit replacing the 
cross-sections for hadrons, muon and tau pairs 
with the ratios $\sigma(\mu\mu)/\sigma(\qqbar)$ and 
$\sigma(\tau\tau)/\sigma(\qqbar)$. The result of this fit is
$1/\alphaem(169.34~\mathrm{GeV}) = 126.8^{+5.0}_{-4.4}$, with a
$\chi^2$ of 8.7 for 15 degrees of freedom. The value is close to
that obtained from the cross-section fit but with somewhat larger errors. 
The difference in $\chi^2$ between the best fit and the assumption
that $\alphaem$ does not run with energy but is fixed at the low
energy limit is $(1.91)^2$. If $\alphaem$ did not run with energy,
the probability of measuring $1/\alphaem$ lower than 137.0359979 by this 
amount would be 2.8\%, thus demonstrating the running of $\alphaem$ from
$(Q^2 = 0)$ to LEP2 energies.
This measurement of $\alphaem$ is independent of low-mass hadronic loops 
and nearly independent of the mass of the Higgs boson and $\alphas$; 
it can be scaled to the mass of the Z, giving 
$1/\alphaem(91.19~\mathrm{GeV}) = 127.8^{+4.5}_{-3.9}$.

\begin{table}[htbp]
\centering
\begin{tabular}{|c||c|c||c|c|}
\hline
\hline
      &\multicolumn{2}{c||}{Fit} &\multicolumn{2}{c|}{Standard Model} \\
\hline
$\sqrt{s}$ (GeV) &$1/\alphaem$ &$\chi^2$/d.o.f. &$1/\alphaem$ 
&$\chi^2$/d.o.f. \\
\hline
133.29   &125.1$^{+8.1}_{-6.4}$  & 2.2/4  & 128.3  & 2.4/5 \\
182.69   &131.2$^{+6.5}_{-5.3}$  & 2.2/4  & 127.9  & 2.6/5 \\
\hline
169.34   &127.8$^{+4.1}_{-3.6}$  &11.7/19 & 128.0  &11.7/20 \\
\hline
\hline
\end{tabular}
\caption[]{Results of fits for $\alphaem$. The first two rows show
 the fits to data at each centre-of-mass energy, the last row the combined
 fit to these energies and measurements at 161 and 
 172~GeV~\cite{bib:OPAL-SM172}. 
 The Standard Model values of $1/\alphaem$, and
 the $\chi^2$ between the measurements and the Standard Model predictions
 are also given for comparison.
}
\label{tab:alphaem1}
\end{table}

\section{Constraints on New Physics}     \label{sec:new_phys}
New physics would be revealed by deviations of the measured data from
Standard Model predictions. The good agreement between data and the
Standard Model places severe constraints on the energy scale of new
phenomena, which are investigated in this section.

For effects arising from the exchange of a new particle with mass 
$\mX$ the contact interaction offers an appropriate framework for
$\mX\gg\sqrt{s}$. Limits on the energy scale $\Lambda$ are presented
for various models. For lower mass ranges, \mbox{$\sqrt{s}$
\raisebox{2pt}{\mbox{$<$}}\makebox[-8pt]{\raisebox{-3pt}{$\sim$}\,} $\
\, \mX < \Lambda$,} propagator and width effects must be taken into
account. The results of a search for heavy particles which couple
to leptons, or to both quarks and leptons, are reported. A new technique
has been developed in which a scan over the complete $s'$ distribution
significantly improves our sensitivity to the $s$-channel process
mediated by particles such as $R$-parity violating sneutrinos.

\subsection{Limits on Four-fermion Contact Interactions}  \label{sec:ci}
A very general framework in which to search for the effect of new
physics is the four-fermion contact interaction. In this 
framework~\cite{bib:Eichten} the Standard Model Lagrangian for 
$\epem\to\ff$ is extended by a term describing a new effective interaction 
with an unknown coupling constant $g$ and an energy scale $\Lambda$:
\begin{eqnarray}\label{eq-contact}
{\cal L}^{\mathrm{contact}} & = 
         & \frac{g^2}{(1 + \delta)\Lambda^2}
         \sum_{i,j=\mathrm{L,R}}\eta_{ij}[\bar{\Pe}_i\gamma^{\mu}{\Pe}_i]
                                       [\bar{\rm f}_j\gamma_{\mu}{\rm f}_j] ,
\end{eqnarray} 
where $\delta = 1$ for $\epem\to\epem$ and $\delta = 0$ otherwise.
Here $\mathrm{e_L} (f_\mathrm{L})$ and $\mathrm{e_R} (f_\mathrm{R})$ 
are chirality projections of electron (fermion) 
spinors, and $\eta_{ij}$ describes the chiral structure of 
the interaction. The parameters $\eta_{ij}$ are free in these models, but
typical values are between $-1$ and $+1$, depending on the type of theory 
assumed~\cite{bib:contacttable}. Here we consider the same set of 
models as in~\cite{bib:OPAL-SM172}.

We have repeated the analysis described in~\cite{bib:OPAL-SM172}, including
the measurements of the angular distributions for the non-radiative
\mbox{$\eetoee$,} $\eetomumu$, $\eetotautau$ processes, the 
non-radiative cross-section for $\eetoqq$, and the measurement of $\Rb$
at 183~GeV presented here. We have also replaced the measurements at
130--136~GeV used in~\cite{bib:OPAL-SM172} with the values from the
combined 1995+1997 data presented here. As before, we use a maximum
likelihood fit in the case of the lepton angular distributions, and
a $\chi^2$ fit for the hadronic and \bbbar\ cross-sections. Radiative
corrections to the lowest order cross-section are taken into account
as described in~\cite{bib:OPAL-SM172}. Limits on the energy scale 
$\Lambda$ are extracted assuming $g^2/4 \pi = 1$.

\begin{figure}
\begin{center}
\epsfxsize=0.86\textwidth
\epsfbox{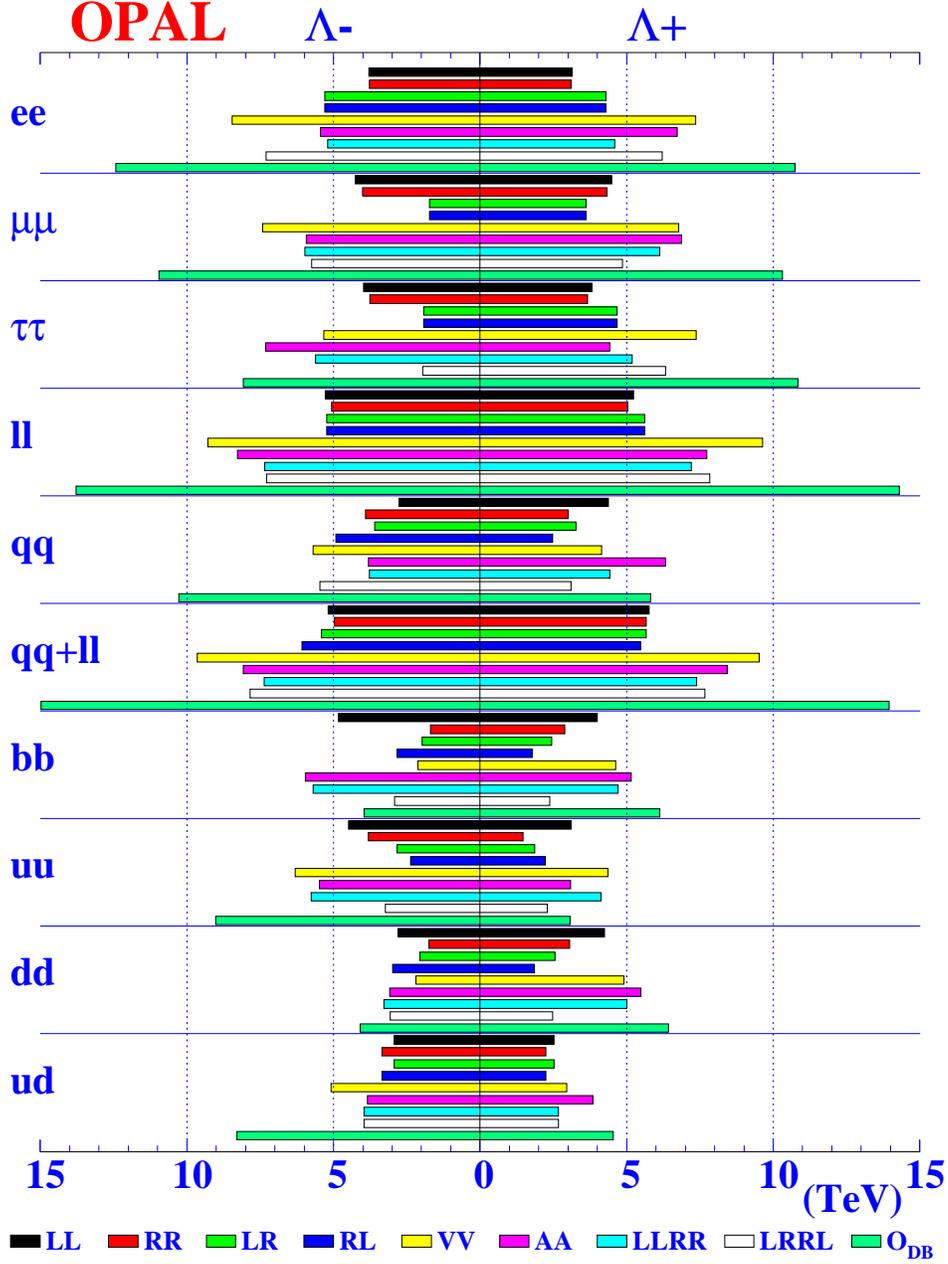}
\caption{95\% confidence level limits on the energy scale $\Lambda$
resulting from the contact interaction fits. For each channel, the
bars from top to bottom indicate the results for models LL to 
$\overline{\cal{O}}_{\mathrm{DB}}$ in the order given in the key.
}
\label{fig:ccres} 
\end{center}
\end{figure}

The results are shown in Table~\ref{tab:ccres} and illustrated
graphically in Fig.~\ref{fig:ccres}; the notation for
the different models is identical to~\cite{bib:OPAL-CI,bib:OPAL-SM172}.
The values for $\bbbar$ are obtained by fitting the cross-sections
for $\bbbar$ production, and there is no requirement on whether or
not the new interaction couples to other flavours. By contrast,
those for up-type quarks and down-type quarks are obtained by fitting 
the hadronic cross-sections assuming the new interaction couples only to one
flavour. The two sets of values $\Lambda_+$ and $\Lambda_-$ shown in 
Table~\ref{tab:ccres} correspond to positive and negative values of 
$\varepsilon = 1/\Lambda^2$ respectively, reflecting the two possible signs 
of $\eta_{ij}$ in equation~\eqref{eq-contact}.
As before, the data are particularly sensitive to the 
VV and AA models; the combined data give limits on $\Lambda$ in the
range 8--10~TeV for these models, roughly 2~TeV higher than
those for 130--172~GeV data alone. For the other models the limits
generally lie in the range 5--8~TeV, approximately 1.5~TeV above those from 
previous data. Those for the $\overline{\cal{O}}_{\mathrm{DB}}$ 
model~\cite{bib:CInew2} are larger (14--15~TeV) because the values of the 
$\eta$ parameters are larger for this model.

\begin{table}[htbp]
\begin{sideways}
\begin{minipage}[b]{\textheight}{\footnotesize
\begin{center}\begin{tabular}{|cc|c|c|c|c|c|c|c|c|c|}
\hline
Channel &      &  LL  &  RR  &  LR  &  RL  &  VV  &  AA  &  
               LL+RR  &LR+RL & $\overline{\cal O}_{\mathrm{DB}}$ \\
        &      &  \scriptsize{ $[\pm1,0,0,0]$}  & 
                  \scriptsize{ $[0,\pm1,0,0]$}  & 
                  \scriptsize{ $[0,0,\pm1,0]$}  &
                  \scriptsize{ $[0,0,0,\pm1]$}  & 
                  \scriptsize{ $[\pm1,\pm1,\pm1,\pm1]$} &
                  \scriptsize{ $[\pm1,\pm1,\mp1,\mp1]$} &
                  \scriptsize{ $[\pm1,\pm1,0,0]$} & 
                  \scriptsize{ $[0,0,\pm1,\pm1]$} &
                  \scriptsize{ $[\pm1,\pm4,\pm2,\pm2]$} \\

 \hline
\epem    &$\epsz$& $ 0.009_{-0.042}^{+0.044}$ & $ 0.009_{-0.043}^{+0.045}$ & 
                   $ 0.006_{-0.021}^{+0.024}$ & $ 0.006_{-0.021}^{+0.024}$ & 
                   $ 0.003_{-0.008}^{+0.008}$ & $-0.003_{-0.015}^{+0.013}$ & 
                   $ 0.004_{-0.022}^{+0.021}$ & $ 0.003_{-0.011}^{+0.011}$ & 
                   $ 0.001_{-0.004}^{+0.004}$ \\
         &$\lamp$& 3.1 & 3.1 & 4.3 & 4.3 & 7.4 & 6.7 & 4.6 & 6.2 &10.8 \\
         &$\lamm$& 3.8 & 3.8 & 5.3 & 5.3 & 8.5 & 5.5 & 5.2 & 7.3 &12.4 \\
 \hline
\mumu    &$\epsz$& $-0.001_{-0.027}^{+0.026}$ & $-0.003_{-0.030}^{+0.029}$ & 
                   $ 0.013_{-0.037}^{+0.034}$ & $ 0.013_{-0.037}^{+0.034}$ & 
                   $ 0.002_{-0.010}^{+0.010}$ & $-0.004_{-0.012}^{+0.013}$ & 
                   $-0.001_{-0.014}^{+0.014}$ & $ 0.007_{-0.019}^{+0.018}$ & 
                   $ 0.001_{-0.005}^{+0.005}$ \\
         &$\lamp$& 4.5 & 4.3 & 3.6 & 3.6 & 6.8 & 6.9 & 6.1 & 4.9 &10.3 \\
         &$\lamm$& 4.3 & 4.0 & 1.7 & 1.7 & 7.4 & 5.9 & 6.0 & 5.7 &11.0 \\
 \hline
\tautau  &$\epsz$& $ 0.005_{-0.035}^{+0.032}$ & $ 0.006_{-0.038}^{+0.036}$ & 
                   $-0.111_{-0.103}^{+0.078}$ & $-0.111_{-0.103}^{+0.078}$ & 
                   $-0.009_{-0.013}^{+0.013}$ & $ 0.018_{-0.016}^{+0.017}$ & 
                   $ 0.002_{-0.017}^{+0.018}$ & $-0.048_{-0.036}^{+0.030}$ & 
                   $-0.004_{-0.006}^{+0.006}$ \\
         &$\lamp$& 3.8 & 3.7 & 4.7 & 4.7 & 7.4 & 4.4 & 5.2 & 6.3 &10.8 \\
         &$\lamm$& 4.0 & 3.8 & 1.9 & 1.9 & 5.3 & 7.3 & 5.6 & 2.0 & 8.1 \\
 \hline
$\lept$  &$\epsz$& $ 0.001_{-0.019}^{+0.018}$ & $ 0.001_{-0.020}^{+0.020}$ & 
                   $-0.003_{-0.017}^{+0.018}$ & $-0.003_{-0.017}^{+0.018}$ & 
                   $ 0.000_{-0.006}^{+0.006}$ & $ 0.001_{-0.008}^{+0.008}$ & 
                   $ 0.000_{-0.010}^{+0.010}$ & $-0.002_{-0.009}^{+0.009}$ & 
                   $ 0.000_{-0.003}^{+0.003}$ \\
         &$\lamp$& 5.2 & 5.0 & 5.6 & 5.6 & 9.6 & 7.7 & 7.2 & 7.8 &14.3 \\
         &$\lamm$& 5.3 & 5.1 & 5.2 & 5.2 & 9.3 & 8.3 & 7.4 & 7.3 &13.8 \\
 \hline
\qqbar   &$\epsz$& $-0.037_{-0.063}^{+0.057}$ & $ 0.023_{-0.056}^{+0.057}$ & 
                   $ 0.007_{-0.054}^{+0.055}$ & $ 0.024_{-0.039}^{+0.105}$ & 
                   $ 0.013_{-0.028}^{+0.030}$ & $-0.019_{-0.035}^{+0.028}$ & 
                   $-0.011_{-0.037}^{+0.040}$ & $ 0.023_{-0.036}^{+0.060}$ & 
                   $ 0.007_{-0.010}^{+0.017}$ \\
         &$\lamp$& 4.4 & 3.0 & 3.3 & 2.5 & 4.1 & 6.3 & 4.4 & 3.1 & 5.8 \\
         &$\lamm$& 2.8 & 3.9 & 3.6 & 4.9 & 5.7 & 3.8 & 3.8 & 5.5 &10.3 \\
 \hline
combined &$\epsz$& $-0.003_{-0.017}^{+0.017}$ & $-0.004_{-0.018}^{+0.018}$ & 
                   $-0.002_{-0.016}^{+0.017}$ & $ 0.002_{-0.015}^{+0.016}$ & 
                   $ 0.000_{-0.006}^{+0.006}$ & $ 0.000_{-0.008}^{+0.007}$ & 
                   $ 0.000_{-0.010}^{+0.009}$ & $ 0.000_{-0.008}^{+0.009}$ & 
                   $ 0.000_{-0.002}^{+0.003}$ \\
         &$\lamp$& 5.8 & 5.7 & 5.7 & 5.5 & 9.5 & 8.4 & 7.4 & 7.7 &14.0 \\
         &$\lamm$& 5.2 & 5.0 & 5.4 & 6.1 & 9.7 & 8.1 & 7.4 & 7.9 &15.0 \\
 \hline
\bbbar   &$\epsz$& $ 0.015_{-0.027}^{+0.025}$ & $ 0.035_{-0.068}^{+0.049}$ & 
                   $-0.154_{-0.059}^{+0.277}$ & $-0.039_{-0.048}^{+0.077}$ & 
                   $ 0.014_{-0.024}^{+0.017}$ & $ 0.010_{-0.018}^{+0.014}$ & 
                   $ 0.011_{-0.019}^{+0.018}$ & $-0.046_{-0.042}^{+0.187}$ & 
                   $-0.046_{-0.010}^{+0.018}$ \\
         &$\lamp$& 4.0 & 2.9 & 2.4 & 1.8 & 4.6 & 5.2 & 4.7 & 2.4 & 6.1 \\
         &$\lamm$& 4.8 & 1.7 & 2.0 & 2.8 & 2.1 & 6.0 & 5.7 & 2.9 & 4.0 \\
 \hline
\uubar   &$\epsz$& $ 0.021_{-0.037}^{+0.039}$ & $ 0.033_{-0.056}^{+0.069}$ & 
                   $ 0.069_{-0.126}^{+0.152}$ & $ 0.001_{-0.112}^{+0.126}$ & 
                   $ 0.010_{-0.018}^{+0.020}$ & $ 0.015_{-0.026}^{+0.034}$ & 
                   $ 0.012_{-0.021}^{+0.023}$ & $ 0.033_{-0.081}^{+0.105}$ & 
                   $ 0.005_{-0.009}^{+0.011}$ \\
         &$\lamp$& 3.1 & 1.5 & 1.9 & 2.2 & 4.4 & 3.1 & 4.1 & 2.3 & 3.1 \\
         &$\lamm$& 4.5 & 3.8 & 2.8 & 2.4 & 6.3 & 5.5 & 5.8 & 3.2 & 9.0 \\
 \hline
\ddbar   &$\epsz$& $-0.023_{-0.047}^{+0.041}$ & $-0.077_{-0.184}^{+0.118}$ & 
                   $-0.040_{-0.126}^{+0.124}$ & $ 0.077_{-0.120}^{+0.149}$ & 
                   $-0.020_{-0.155}^{+0.034}$ & $-0.015_{-0.034}^{+0.026}$ & 
                   $-0.016_{-0.035}^{+0.029}$ & $ 0.035_{-0.089}^{+0.081}$ & 
                   $-0.015_{-0.031}^{+0.026}$ \\
         &$\lamp$& 4.3 & 3.1 & 2.6 & 1.8 & 4.9 & 5.5 & 5.0 & 2.5 & 6.4 \\
         &$\lamm$& 2.8 & 1.8 & 2.1 & 3.0 & 2.2 & 3.1 & 3.3 & 3.1 & 4.1 \\
 \hline
\ud      &$\epsz$& $ 0.020_{-0.087}^{+0.087}$ & $ 0.054_{-0.093}^{+0.098}$ & 
                   $ 0.020_{-0.087}^{+0.087}$ & $ 0.054_{-0.093}^{+0.098}$ & 
                   $ 0.029_{-0.044}^{+0.064}$ & $ 0.001_{-0.043}^{+0.042}$ & 
                   $ 0.037_{-0.066}^{+0.069}$ & $ 0.037_{-0.066}^{+0.069}$ & 
                   $ 0.009_{-0.015}^{+0.030}$ \\
         &$\lamp$& 2.5 & 2.2 & 2.5 & 2.2 & 3.0 & 3.9 & 2.7 & 2.7 & 4.5 \\
         &$\lamm$& 2.9 & 3.4 & 2.9 & 3.4 & 5.1 & 3.9 & 4.0 & 4.0 & 8.3 \\
 \hline
\end{tabular}\end{center}}
\caption[foo]{\label{tab:ccres}
  Results of the contact interaction fits to the angular distributions for
  non-radiative $\eetoee$, $\eetomumu$, $\eetotautau$, the cross-sections
  for $\eetoqq$ and the measurements of $\Rb$. Results at centre-of-mass 
  energies of 161--172~GeV~\cite{bib:OPAL-SM172} are also included.
  The combined results include all leptonic angular distributions and the 
  hadronic cross-sections. The numbers in square brackets are the values of
  [$\eta_{\mathrm{LL}}$,$\eta_{\mathrm{RR}}$,$\eta_{\mathrm{LR}}$,
  $\!\eta_{\mathrm{RL}}$] which define the models.
  $\epsz$ is the fitted value of $\varepsilon = 1/\Lambda^{2}$,
  $\Lambda_{\pm}$ are the 95\% confidence level limits.
  The units of $\Lambda$ are TeV, those of $\epsz$ are $\mathrm{TeV}^{-2}$.
}
\end{minipage}
\end{sideways}
\end{table}
\subsection{Limits on Heavy Particles}           \label{sec:lq}
In this section we present the results of a search under
the explicit assumption that the new phenomena are due to a heavy
particle which couples to leptons, or to quarks and leptons. The
sensitivity of the OPAL data to these phenomena is tested separately
for the hadronic and leptonic events. Although we use specific
particles in the analyses presented below, the results are generally 
applicable for any heavy particle with similar properties.

\subsubsection{Particles Coupling to Quarks and Leptons}
Examples of new heavy particles which could contribute to the
hadronic cross-section are leptoquarks~\cite{bib:lqbas} or squarks in
supersymmetric theories with $R$-parity violation~\cite{bib:rpvbas}.  
Beyond the kinematic limit for direct production, such a new particle might 
be seen through a change of the total cross-section in the process  
$\epem\to\qqbar$ via a $t$-channel exchange diagram. The classification   
of the various allowed leptoquark states is given in~\cite{bib:lqsig}.

To search for new heavy particles coupling to quarks and leptons we 
perform a $\chi^2$ fit between measured non-radiative hadronic
cross-sections and model predictions, as described in detail
in~\cite{bib:OPAL-SM172}. The hadronic cross-sections used are the values 
at 183~GeV and 130--136~GeV presented here, together with those at 161 and 
172~GeV presented in~\cite{bib:OPAL-SM172}. Correlations between these
measurements caused by common systematic errors are taken into account.
We also perform fits to the \bbbar\ cross-sections, considering all
possible leptoquark couplings to the b quark.
The predicted cross-section for $\epem \rightarrow \qqbar$ including 
$t$-channel exchange of a leptoquark is calculated in~\cite{bib:lqsig}. 
Electroweak corrections are included as for the contact interaction fits, and
Standard Model cross-sections are calculated using ZFITTER.

\begin{figure}
\begin{center}
\epsfxsize=0.83\textwidth
\epsfbox[50 90 545 792]{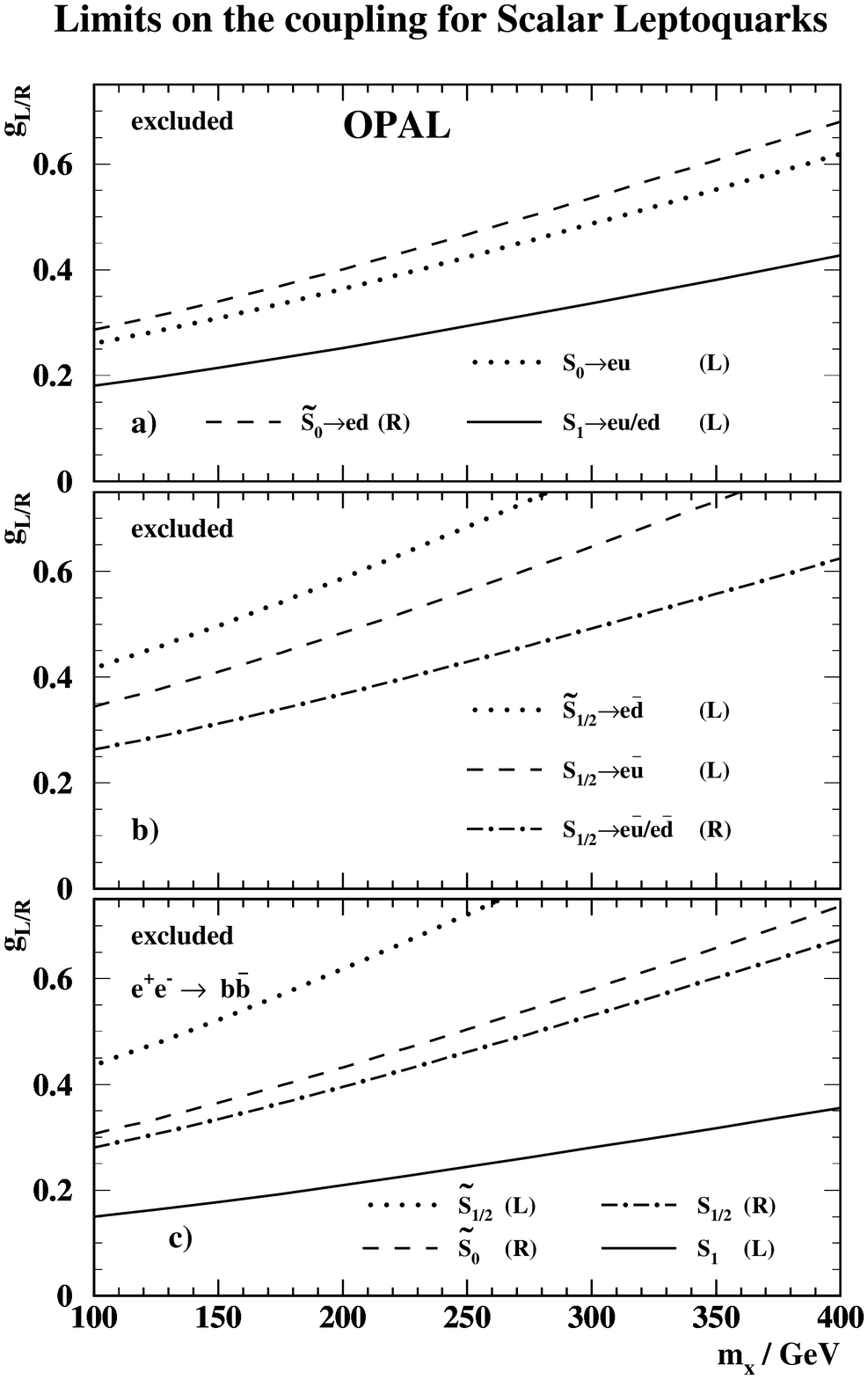}
\caption{95\% confidence exclusion limits on $\mathrm{g_{L}}$ or 
$\mathrm{g_{R}}$ as a function of $\protect\mX$, for various possible 
scalar leptoquarks. (a) and (b) show limits on leptoquarks coupling 
to a single quark family, derived from the hadronic cross-sections.
(c) shows limits on leptoquarks coupling to b quarks only, derived from
the \Pb\Pab\ cross-sections. The excluded regions are above the curves
in all cases.
The letter in parentheses after the different leptoquark types indicates
the chirality of the lepton involved in the interaction. The limits on
the $\mathrm{S}_{0}$ and $\tilde{\mathrm{S}}_{1/2}$ leptoquarks can 
be interpreted as limits on $R$-parity violating 
$\tilde{\mathrm{d}}_{\mathrm{R}}$ and $\tilde{\mathrm{u}}_{\mathrm{L}}$ 
squarks respectively.
}
\label{fig:limits_scalar} 
\end{center}
\end{figure}

\begin{figure}
\begin{center}
\epsfxsize=0.83\textwidth
\epsfbox[50 90 545 792]{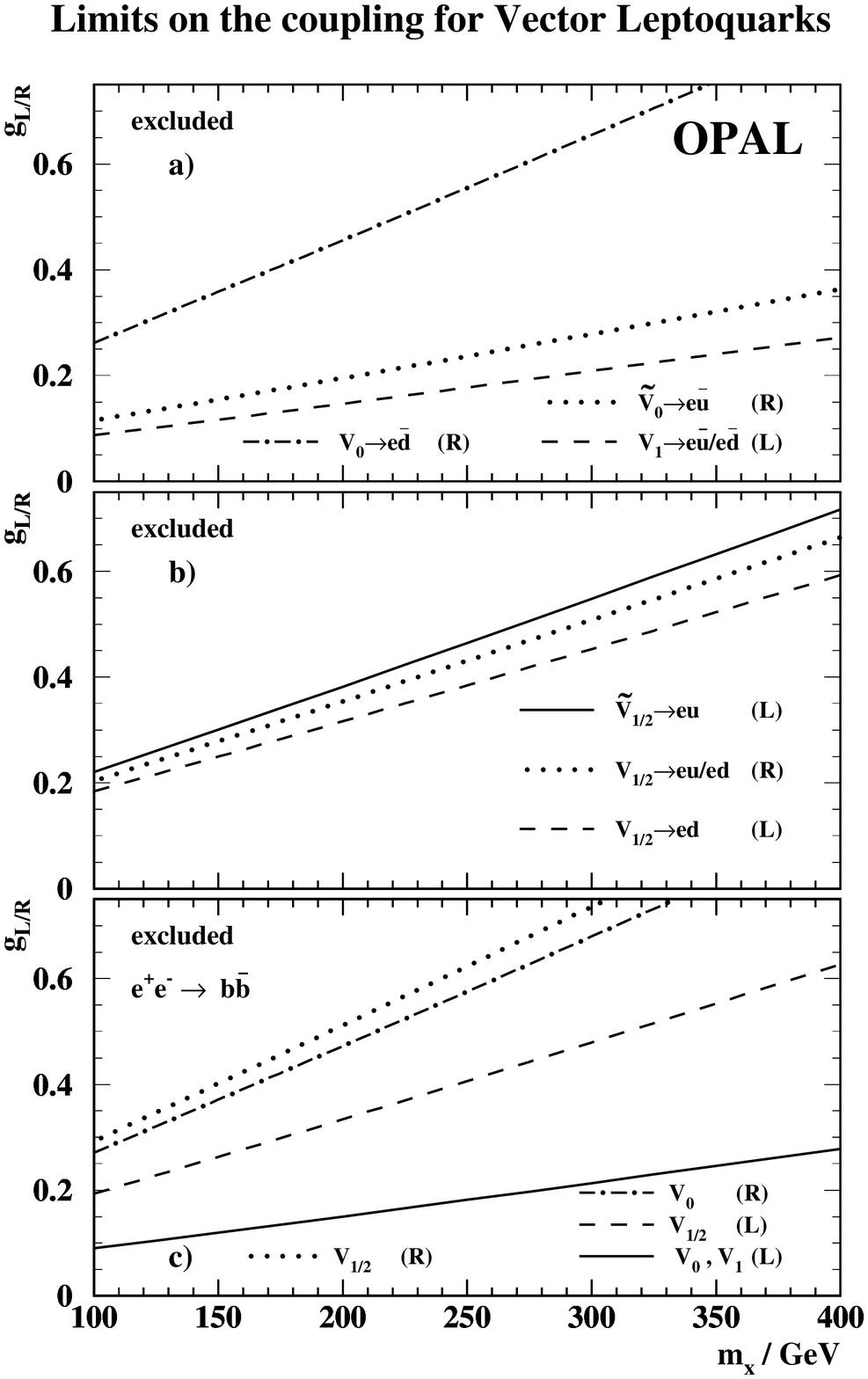}
\caption{95\% confidence exclusion limits on $\mathrm{g_{L}}$ or 
$\mathrm{g_{R}}$ as a function of $\protect\mX$, for various possible 
vector leptoquarks. (a) and (b) show limits on leptoquarks coupling 
to a single quark family, derived from the hadronic cross-sections.
(c) shows limits on leptoquarks coupling to b quarks only, derived from
the \Pb\Pab\ cross-sections. The excluded regions are above the curves
in all cases.
The letter in parentheses after the different leptoquark types indicates
the chirality of the lepton involved in the interaction. 
}
\label{fig:limits_vector} 
\end{center}
\end{figure}

Figure~\ref{fig:limits_scalar} shows the 95\%~confidence limits obtained
as a function of the mass $\mX$ and the coupling constant $\mathrm{g_L}$ 
or $\mathrm{g_R}$ of the new particle, for scalar states. Results 
for the vector leptoquark states are shown in Fig.~\ref{fig:limits_vector}. 
We do not show limits on the $\rm S_0$ ($\rm V_0$) leptoquark with coupling 
$\rm g_R$ ($\rm g_L$) because the effect of these particles on the hadronic 
cross-section at these energies is too small. For the states shown, the 
limits on the coupling, derived from hadronic cross-section measurements,
typically lie in the range 0.15--0.6 for a mass of 200~GeV. Inclusion of 
data at 183~GeV has lowered these limits by about 20\% compared with those 
obtained from lower energy data~\cite{bib:OPAL-SM172,bib:L3-newphys}.
The limits derived from \bbbar\ cross-sections are generally somewhat more
stringent than those obtained from the hadronic cross-sections.

As can be seen in the figures our analysis is sensitive to
leptoquark masses much higher than the beam energy. 
Direct searches at the Tevatron can exclude
scalar  and vector leptoquarks with Yukawa couplings down to 
${\cal O}(10^{-7})$ up to masses of $\approx 225$~GeV \cite{bib:tevlim}.
Our limits extend this excluded region for large couplings. 

\subsubsection{Particles Coupling to Leptons}
The cross-section for the production of two leptons can also be 
affected by the presence of new heavy particles which couple to 
leptons. Three cases can be distiguished: the new particles might 
contribute to the measured cross-section via the $s$-channel, via a 
$t$-channel exchange, or both channels may be present. In order to test the 
sensitivity of the  OPAL data to such interactions, we have chosen to
use sneutrinos with $R$-parity violating couplings as an example. 
Their coupling to leptons is given by the term 
$\lambda_{ijk}L^i_{\rm L}L^j_{\rm L}\overline{E}^k_{\rm R}$  
of the superpotential~\cite{bib:rpvsup}, where the indices $i,j,k$
denote the family of the particles involved, $L^i_{\rm L}$ and $L^j_{\rm L}$ 
are the SU(2) doublet lepton superfields and $\overline{E}^k_{\rm R}$ denotes 
an antilepton singlet superfield. The couplings $\lambda_{ijk}$ are 
non-vanishing only for $i < j$, so at least two different generations of 
leptons are coupled in purely leptonic vertices.
We consider three typical cases in our analysis in order to illustrate
the sensitivity of each of the leptonic channels: 
\begin{itemize}
 \item the presence of a $\tilde \nu_e$ with coupling $\lambda_{131}$,
       giving rise to a change in the \tautau\ cross-section due to 
       $t$-channel exchange;
 \item the presence of a $\snu_\tau$ which interacts via the 
       coupling $\lambda_{131}$ giving rise to a change in the \epem\
       cross-section via an $s$-channel and a $t$-channel process;
       the limits obtained for this case could equally apply to a $\snu_\mu$
       interacting via the coupling $\lambda_{121}$;
 \item a $\tilde \nu_\tau$ with the couplings $\lambda_{131}$ and  
       $\lambda_{232}$ both different from zero. In the analysis both 
       couplings are assumed to be of equal size\footnote{Both 
       couplings violate the same lepton flavours so that this scenario 
       is compatible with the experimental observation of lepton 
       number conservation.}. Such a scenario gives rise to a modified \mumu\
       cross-section due to an $s$-channel exchange of the sneutrino. 
\end{itemize}
To calculate the differential cross-sections for these processes we use the 
formulae in~\cite{bib:rpvsnu}, taking radiative corrections into account as
in the contact interaction analysis.

In the case of the \tautau\ cross-section only the $t$-channel 
exchange of a $\tilde \nu_\mathrm{e}$ is involved. We perform a maximum 
likelihood fit which compares the model prediction with the observed
number of events having $s'/s > 0.7225$. Systematic errors are 
taken into account by multiplying the likelihood by a factor $(1+r)$ where
$r$ is distributed according to a Gaussian with a width given by the 
systematic error on the data sample. This is the method used to
extract limits on four-fermion contact interactions in this paper
and in~\cite{bib:OPAL-CI,bib:OPAL-SM172}. For each mass, the 95\% confidence 
level limit on the coupling is obtained as the value of $\lambda$ 
corresponding to a change in negative log likelihood of 1.92 with respect 
to the minimum.

The cases where an $s$-channel diagram is involved are especially
interesting. A narrow peak in the cross-section when the centre-of-mass
energy equals the mass of the sneutrino is then predicted, since the 
sneutrinos are expected to have a small width 
($\lapproxeq$ 1~GeV~\cite{bib:rpvsnu}). In order 
to improve the sensitivity for sneutrino masses which lie between the 
centre-of-mass energies of LEP, a fit method has been developed which
scans the complete $s'$-distribution of the recorded data. The
$\sqrt{s'}$ distributions at 183~GeV are shown in Fig.~\ref{fig:sp};
a sneutrino of mass $m$ would be expected to produce a peak in the 
$\sqrt{s'}$ distribution at $\sqrt{s'} \simeq m$ for a process receiving 
a contribution from $s$-channel sneutrino production.
We calculate limits on the coupling as a function of sneutrino mass
as follows. For each mass $m$, we consider a window of width $\pm$2.5~GeV
around $\sqrt{s'}=m$, and calculate the number of signal and Standard
Model background events expected in this interval. We assume the efficiency
for detecting signal events (within the $\absct$ acceptance of each
channel) is the same as that for Standard Model
events at the same $s'$. The size of the window is chosen 
to be such that at least 60\% of events of mass $m$ are reconstructed
within the window. In this interval a maximum likelihood fit is performed 
which compares the model prediction with the measured data. Systematic 
errors are considered in the same way as for the $t$-channel case. 

In the case of the \epem\ cross-section the presence of both an $s$-channel 
and a $t$-channel diagram is assumed, and we consider two contributions
to the likelihood function: one arising from the $\sqrt{s'}$ interval
around $m$ and one from the $s'/s > 0.7225$ region. In cases where these
intervals overlap only the contribution from the interval around $m$
was considered, as the $s$-channel diagram is expected to dominate.
For this analysis, we use electron pairs with $\absctem < 0.7$, 
with no cut on acollinearity. The Standard Model prediction in each $s'$ 
window is calculated using Monte Carlo events generated with 
BHWIDE~\cite{bib:bhwide}. Systematic errors on the Standard Model prediction 
are assessed by reweighting the Monte Carlo events to the acollinearity 
distribution predicted by ALIBABA.

At large values of $\lambda$ the limits can be improved by including
measurements of forward-backward asymmetry in the fit. This is
achieved by adding a term to the negative log likelihood of the form
\[
-\ln P = \sum_{i} \ln \Delta A_{i} + 
 \frac{(A^{\mathrm{th}}_{i} - A^{\mathrm{meas}}_{i})^2}
      {2 (\Delta A_{i})^2}
\]
where the sum runs over the measurements at different energies.
$A^{\mathrm{th}}_{i}$ is the theoretical prediction of the asymmetry,
including the effects of new physics, $A^{\mathrm{meas}}_{i}$ is the
measured value and $\Delta A_{i}$ the error on the measurement. We
have calculated limits both including and excluding measurements of
asymmetry for non-radiative events. While the former are valid for 
the exchange of a scalar particle with a width less than or equal to 1~GeV, 
the latter 
are less model dependent and could be reinterpreted in terms of the
exchange of a vector particle with appropriate modifications to the
coupling constant. For the \epem\ and \mumu\ channels, the inclusion
of asymmetry values has a very small effect on the limits, therefore
we present only limits excluding asymmetry measurements in order not to
lose generality. In the case of the \tautau\ channel the limits on
$\lambda$ are larger and we present results both including and excluding the 
asymmetry measurements.

In all cases, we include in the fits the data at 183 and 130--136~GeV
presented here and the data at 161 and 172~GeV presented 
in~\cite{bib:OPAL-SM172}. In the analysis of the $\tautau$ channel
we use the measured non-radiative cross-sections and asymmetries,
whereas for the $\mumu$ and $\epem$ channels we use the full
$\sqrt{s'}$ distributions as described above. We present limits on the 
coupling as a function of sneutrino mass for each of the three cases.

\begin{figure}
\begin{center}
\epsfxsize=\textwidth
\epsfbox{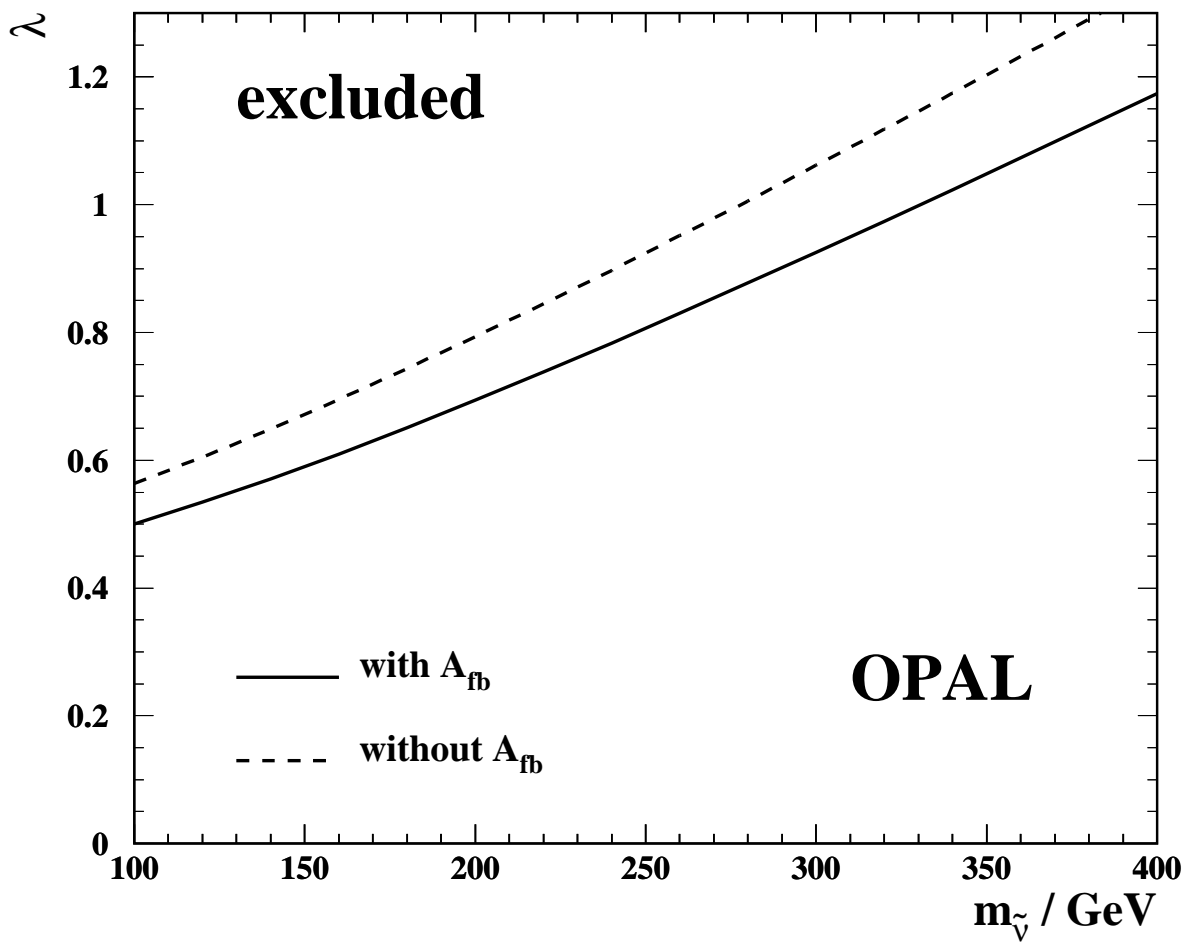}
\caption{95\% confidence exclusion limit on $\lambda_{131}$ as a function 
 of sneutrino mass $\protect m_{\snu}$, derived from \tautau\ cross-section
 and asymmetry data. The region above the solid line is excluded.
 The dashed line shows the limit determined from cross-section data alone.
}
\label{fig:tau_limits} 
\end{center}
\end{figure}

\begin{figure}
\begin{center}
\epsfxsize=\textwidth
\epsfbox{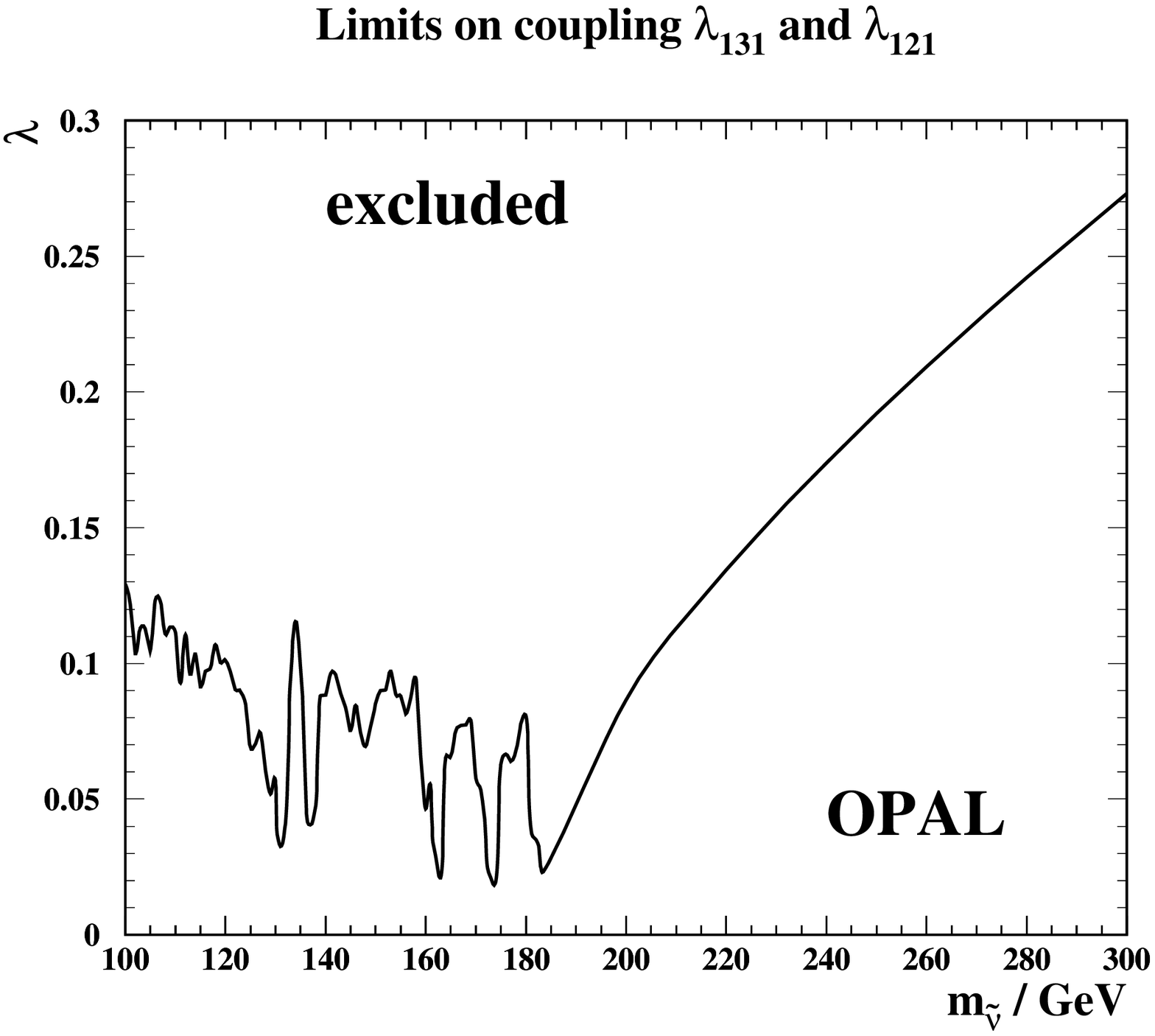}
\caption{95\% confidence exclusion limits on $\protect\lambda_{131}$ 
 (or $\protect\lambda_{121}$) as a function of sneutrino mass 
 $\protect m_{\snu}$, derived from \epem\ $s'$ distributions. The region 
 above the solid line is excluded.
}
\label{fig:ee_limits} 
\end{center}
\end{figure}

\begin{figure}
\begin{center}
\epsfxsize=\textwidth
\epsfbox{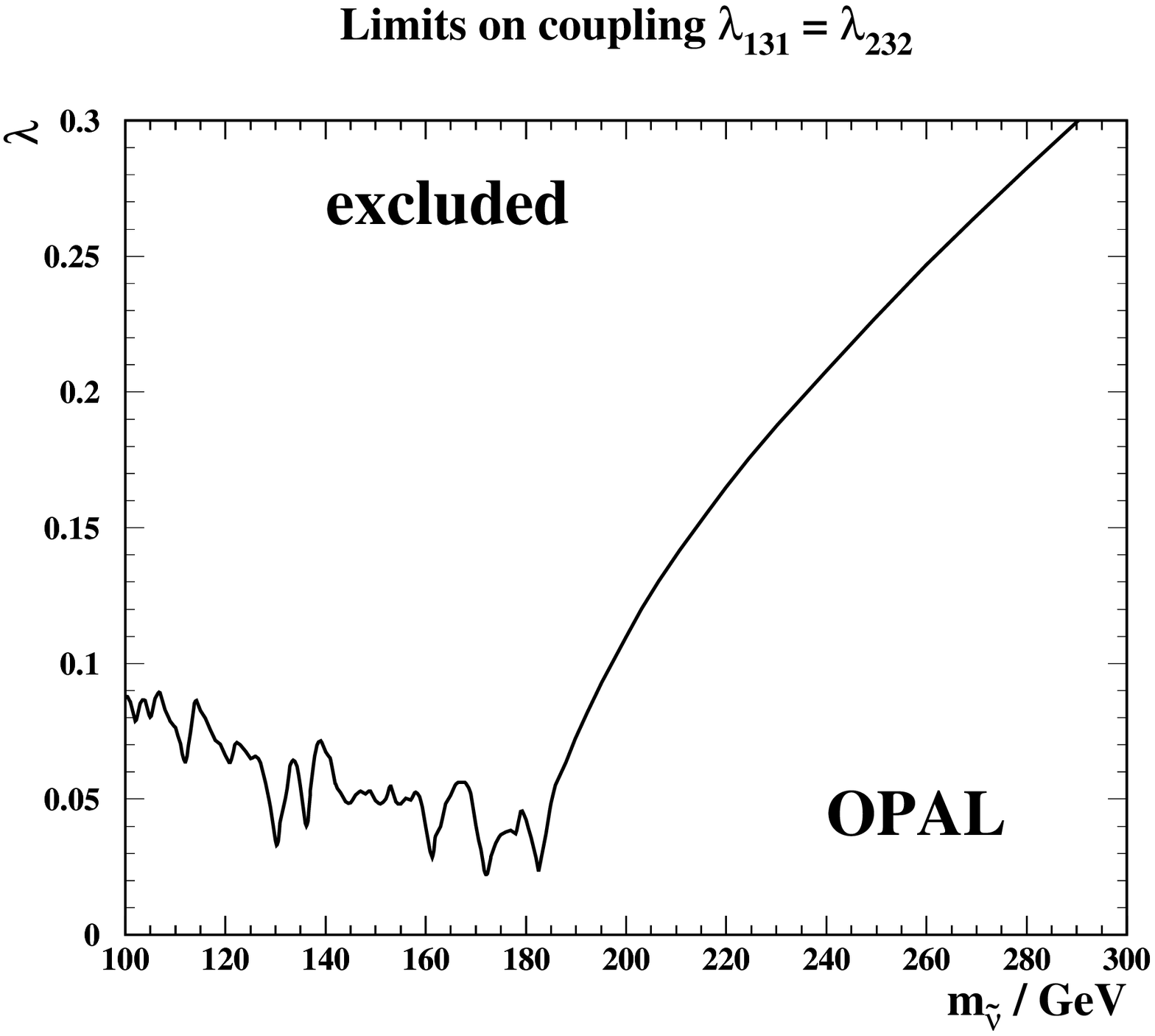}
\caption{95\% confidence exclusion limit on $\protect\lambda_{131} = 
 \lambda_{232}$ as a function of sneutrino mass $\protect m_{\snu}$, derived 
 from \mumu\ $s'$ distributions. The region above the solid
 line is excluded.
}
\label{fig:mu_limits} 
\end{center}
\end{figure}

The limits on $\lambda_{131}$ derived from the \tautau\ data are shown in 
Fig.~\ref{fig:tau_limits}. As only a $t$-channel contribution is involved,
the limit on $\lambda$ varies smoothly with the $\snu$ mass, from about 
0.5 at 100~GeV to 1.2 at 400~GeV. The inclusion of asymmetry data 
improves the limit by about 0.1. The limits on $\lambda_{131}$ derived from 
the \epem\ data are shown in Fig.~\ref{fig:ee_limits}. They are in the range 
\mbox{0.02 -- 0.1} for $100 < m < 200$~GeV, rising to 0.27 at 300~GeV.
Figure~\ref{fig:mu_limits} shows limits on $\lambda_{131} = \lambda_{232}$ 
derived from the \mumu\ data. These are in the range 0.02 -- 0.1 for 
$100 < m < 200$~GeV, rising to 0.3 at 300~GeV. By introducing the $s'$ 
distributions in our fits we find the sensitivity in the \epem\ and
\mumu\ channels has been improved for masses between the LEP 
centre-of-mass energy points. The fine structure in the region $m <$ 200~GeV 
results from fluctuations in the $s'$ distributions.

Direct searches~\cite{bib:ALEPH-rpv} for sneutrinos with $R$-parity 
violating couplings can conservatively exclude a $\snu_\tau$ with mass 
less than 49~GeV and a $\snu_e$ with mass less than 72~GeV. Our results
place limits on the couplings for higher masses.

\section{Conclusions}    \label{sec:sum}
We have presented new measurements of cross-sections and asymmetries for
hadron and lepton pair production in $\epem$ collisions at centre-of-mass
energies of 183 and 130--136~GeV, and combined the 130--136~GeV results
with our measurements from earlier data~\cite{bib:OPAL-SM172}. The
results, for both inclusive fermion-pair production and for non-radiative
events, are in good agreement with Standard Model expectations. From
these and earlier measurements we derive a value for the electromagnetic
coupling constant $1/\alphaem (169.34~\mathrm{GeV}) = 127.8^{+4.1}_{-3.6}$.

The measurements have been used to improve existing limits on new
physics. In the context of a four-fermion contact interaction we
have improved the limits on the energy scale $\Lambda$ from typically
2--7~TeV to 2--10~TeV, assuming $g^2/4\pi = 1$. In explicit
searches for new particles coupling to quarks and leptons we obtain
limits on the coupling in the range 0.15--0.6 for a mass of about
200~GeV, some 20\% lower than those obtained from data at centre-of-mass
energies up to 172~GeV. 
We have also presented limits on new particles such as sneutrinos in
supersymmetric theories with $R$-parity violation which couple to
leptons only. Sensitivity to sneutrino masses between the centre-of-mass
energy points of LEP has been improved by using a complete scan of the
$s'$ distribution for processes involving an $s$-channel diagram. In
these cases, limits on the couplings in the range 0.02 -- 0.1 are obtained 
for $100 < m < 200$~GeV.

\section*{Acknowledgements}
We thank R.~R\"{u}ckl for helpful discussions on setting limits on
$R$-parity violating sneutrino couplings.

We particularly wish to thank the SL Division for the efficient operation
of the LEP accelerator at all energies
 and for their continuing close cooperation with
our experimental group.  We thank our colleagues from CEA, DAPNIA/SPP,
CE-Saclay for their efforts over the years on the time-of-flight and trigger
systems which we continue to use.  In addition to the support staff at our own
institutions we are pleased to acknowledge the  \\
Department of Energy, USA, \\
National Science Foundation, USA, \\
Particle Physics and Astronomy Research Council, UK, \\
Natural Sciences and Engineering Research Council, Canada, \\
Israel Science Foundation, administered by the Israel
Academy of Science and Humanities, \\
Minerva Gesellschaft, \\
Benoziyo Center for High Energy Physics,\\
Japanese Ministry of Education, Science and Culture (the
Monbusho) and a grant under the Monbusho International
Science Research Program,\\
German Israeli Bi-national Science Foundation (GIF), \\
Bundesministerium f\"ur Bildung, Wissenschaft,
Forschung und Technologie, Germany, \\
National Research Council of Canada, \\
Research Corporation, USA,\\
Hungarian Foundation for Scientific Research, OTKA T-016660, 
T023793 and OTKA F-023259.\\


%
\clearpage
%
%
%
%
%
%
%
%
%
%
%

\end{document}